\documentclass[]{jfm}

\usepackage{graphicx}
\usepackage{newtxtext}
\usepackage{newtxmath}
\usepackage{natbib}
\usepackage{hyperref}
\hypersetup{
    colorlinks = true,
    urlcolor   = blue,
    citecolor  = black,
}

\usepackage{subfig}
 \usepackage[abs]{overpic}
\usepackage{placeins}
\usepackage{tikz}
\usepackage{bm}

\title{Lagrangian model for passive scalar gradients in turbulence}

\author{Xiaolong Zhang\aff{1},
  Maurizio Carbone\aff{2,3}
 \and Andrew D. Bragg\aff{1}
 \corresp{\email{andrew.bragg@duke.edu}}}

\affiliation{\aff{1}Department of Civil and Environmental Engineering, Duke University, Durham, NC 27708, USA
\aff{2}Max Planck Institute for Dynamics and Self-Organization, Am Faßberg 17, 37077 Göttingen, Germany
\aff{3}Theoretical Physics I, University of Bayreuth, Universitätsstr. 30, 95447 Bayreuth, Germany}

\begin{document}
\maketitle

\begin{abstract}
The equation for the fluid velocity gradient along a Lagrangian trajectory immediately follows from the Navier-Stokes equation. However, such an equation involves two terms that cannot be determined from the velocity gradient along the chosen Lagrangian path: the pressure Hessian and the viscous Laplacian. A recent model handles these unclosed terms using a multi-level version of the recent deformation of Gaussian fields (RDGF) closure (Johnson \& Meneveau, Phys.~Rev.~Fluids, 2017). This model is in remarkable agreement with DNS data and works for arbitrary Taylor Reynolds numbers $\Rey_\lambda$.
Inspired by this, we develop a Lagrangian model for passive scalar gradients in isotropic turbulence. The equation for passive scalar gradients also involves an unclosed term in the Lagrangian frame, namely the scalar gradient diffusion term, which we model using the RDGF approach. However, comparisons of the statistics obtained from this model with direct numerical simulation (DNS) data reveal substantial errors due to erroneously large fluctuations generated by the model. We address this defect by incorporating into the closure approximation information regarding the scalar gradient production along the local trajectory history of the particle. This modified model makes predictions for the scalar gradients, their production rates, and alignments with the strain-rate eigenvectors that are in very good agreement with DNS data. However, while the model yields valid predictions up to around $\Rey_\lambda\approx 500$, beyond this, the model breaks down.
\end{abstract}


\section{\label{sec:level1}Introduction}

 Scalar transport in turbulence plays significant roles in various practical applications, ranging from geophysical and environmental problems such as the advection and dispersion of pollutants in the atmosphere \citep{nironi2015dispersion,mazzitelli2012active}, mixing of nutrients in the ocean \citep{bhamidipati2020turbulent, chor2020mixing}, and chemical reactions in industrial flows \citep{dimotakis2005turbulent, hill1976homogeneous}. Scalar transport is also important from the perspective of fundamental turbulence research, with previous studies showing that the scalar field is a sensitive detector of the structures in turbulent flows, such that their study has yielded insights into the physics of turbulent flows themselves \citep{tong1994passive}. The scalar field may feedback on the velocity field under certain conditions, such as for stratified turbulence or thermally driven turbulence \citep{zhang2022analysis,lohse2010small}. The focus of this paper is, however, on the case of passive scalars.

The transport of scalars in turbulent flows is challenging to understand both because of the complexity of the underlying turbulent flow that advects, stretches and compresses the scalar field, and also because of the molecular diffusion to which it is subject that leads to non-trivial differences compared to the transport of fluid particles \citep{ottino1989kinematics,warhaft2000passive}. Indeed, while it has often been assumed that the statistical properties of passive scalars in turbulent flows should reflect the analogous properties of the underlying velocity field, e.g. a similarity in the statistical distribution of the turbulent kinetic energy and scalar dissipation rates, this is not in general the case. For example, it has been found that Kolmogorov's hypothesis of local isotropy of the small scales of a turbulent flow is strongly violated when applied to passive scalars, with both experiments and direct numerical simulations (DNS) finding that the skewness of the scalar derivative remains of the order of unity when a large-scale mean scalar gradient is imposed, while it would be zero for a locally isotropic flow \citep{sreenivasan1977skewness,sreenivasan1991local,pumir1994numerical,mestayer1976local}. This strong violation of small-scale isotropy of the scalar field is often attributed to the presence of ramp-cliff structures in the scalar field through which large and small scales of the scalar field are directly connected \citep{buaria2021small,shraiman2000scalar}. It has also been found that intermittency in the scalar field is even stronger than that for the velocity field \citep{watanabe2004statistics}. Indeed, even for a stochastic model where the velocity field has Gaussian statistics, the scalar field has non-Gaussian statistics \citep{tong1994passive, kraichnan1994anomalous,falkovich01}. Another profound difference is that while the velocity field exhibits a dissipation anomaly (i.e.~the averaged turbulent kinetic energy dissipation rate is independent of viscosity for high Reynolds numbers), the scalar field does not, with the scalar dissipation rate decreasing as $\sim 1/\log(Sc)$ as the Schmidt number $Sc$ is increased \citep{buaria2021turbulence}. Interestingly, however, recent DNS results have shown that while the expected correspondence between the analogous velocity and scalar statistics is not observed for $Sc\leq 1$, it is recovered for sufficiently large $Sc$ (even for $Sc=7$) where a viscous-convective sub-range emerges in the scalar field \citep{Shete22}.

In general, the properties of a passive scalar field depend upon both the Reynolds number $\Rey$ and Schmidt number $Sc$, and in many practical problems, both $\Rey$ and $Sc$ are large. For $Sc>1$, the smallest scale (in a mean-field sense) in the scalar field is thought to be the Batchelor scale \citep{batchelor1959small} $\eta_B=Sc^{-1/2}\eta$, where $\eta$ is the Kolmogorov length scale, and therefore resolving flows with high $\Rey$ and $Sc$ is very challenging using DNS (as well as experiments) due to the spatial and temporal resolution constraints. For problems where the small-scale properties of the scalar field are important, large eddy simulations are not helpful. It is therefore highly desirable to develop models for the small scales of the scalar field that are capable of handling large ranges of $\Rey$ and $Sc$, as well as being computationally efficient. One possibility is to develop Lagrangian models for the scalar gradients in turbulent flows, inspired by the corresponding models for the velocity gradient that have been highly successful both in terms of making predictions and leading to new insights into the small-scale dynamics of turbulent flows \citep{meneveau2011lagrangian}. Analogous models for scalars could be used to explore and understand the small-scale dynamics of scalar fields at $\Rey$ and $Sc$ that are currently far out of reach using either DNS or experiments.

Lagrangian models for the velocity gradients are derived from the Navier-Stokes equations, but they require modelling/approximations for the pressure Hessian and viscous terms which are unclosed in the reference frame of a single fluid particle trajectory. Various models have been proposed, including the restricted Euler model \citep{vieillefosse1982local}, the tetrad model \citep{chertkov1999lagrangian}, the recent fluid deformation model \citep{chevillard2006lagrangian}, as well as closures based on random Gaussian fields \citep{wilczek2014pressure, johnson2016closure} and, more generally, on tensor function representation of the unclosed terms \citep{Leppin2020}. With the exception of the restricted Euler model, these models predict the steady-state, non-trivial properties of the velocity gradients in turbulent flows, including the preferential alignment of the vorticity with the intermediate strain-rate eigenvector, intermittency, and the multifractal scaling of the moments of the velocity gradients. Unfortunately, most of the models are only capable of predicting flows with low to moderate $\Rey$.
Recently, the multi-level recent deformation of Gaussian fields (ML-RDGF) model has been developed, which was shown to predict DNS data accurately for arbitrary Reynolds numbers \citep{johnson2017turbulence}.

In the equation for a passive scalar gradient along a fluid particle trajectory, the scalar gradient diffusion term is unclosed. A closed model for scalar gradients in turbulent flows was previously derived based on a simple linear relaxation model for the scalar gradient diffusion \citep{martin2005joint}, with the velocity gradient in the equation specified using a restricted Euler model that was modified to include a linear damping term to model viscous effects \citep{martin1998dynamics}. Although the model showed general qualitative agreement with DNS data, there were significant quantitative inaccuracies for a number of key quantities, including significant errors in the predictions for the alignments of the scalar gradients with the strain-rate eigenvectors, and significant underprediction of large fluctuations of the scalar gradients. More advanced models also have been proposed that use the recent fluid deformation approximation \citep{chevillard2006lagrangian} to model the scalar gradient diffusion term \citep{hater2011lagrangian, gonzalez2009kinematic}. These models also yielded qualitatively reasonable predictions, but suffered from quantitative inaccuracies and can only make predictions for relatively low Reynolds numbers due to their use of the recent fluid deformation approximation \citep{chevillard2006lagrangian}. Our work significantly advances these models in two ways.
First, the velocity gradients will be specified using the much more sophisticated ML-RDGF model that also allows for predictions to be made at arbitrarily large $\Rey$. Second, the recent deformation of Gaussian fields closure \citep{johnson2016closure} will be used to provide a more sophisticated closure for the scalar gradient diffusion term. This closure leads to nonlinear terms that can regulate the growth of the scalar gradients in regions of intense stretching, which can be vital to preventing blow-ups from occurring in the model. The closure also incorporates the effect of $Sc$ on the scalar gradient dynamics. In this way, the model is capable of predicting the effect of both $\Rey$ and $Sc$ on the scalar gradient dynamics.

\section{\label{sec:level2_mod_scal}Model for scalar gradients}

\subsection{\label{sec:level3_eq}Governing equations for instantaneous and characteristic variables}

For an incompressible, Newtonian fluid, the equations governing the evolution of the fluid velocity gradient $\boldsymbol{A}\equiv\boldsymbol{\nabla u}$ and scalar gradient $\boldsymbol{B}\equiv\boldsymbol{\nabla}\phi$ are
\begin{align}
D_t\boldsymbol{A}&=-\boldsymbol{A\cdot A}-\boldsymbol{H}+\nu\nabla^2\boldsymbol{A}+\boldsymbol{F_A},\label{Aeq}\\
D_t\boldsymbol{B}&=-\boldsymbol{A}^\top\boldsymbol{\cdot B}+\kappa\nabla^2\boldsymbol{B}+\boldsymbol{F_B},\label{Beq}
\end{align}

where $D_t\equiv \partial_t +\boldsymbol{u\cdot \nabla}$ is the Lagrangian derivative, $\boldsymbol{H}\equiv \boldsymbol{\nabla\nabla}p$ is the pressure Hessian, $p$ is the fluid pressure (normalized by the fluid density), $\nu$ is the fluid kinematic viscosity, $\kappa$ is the scalar diffusivity, and $\boldsymbol{F_A}, \boldsymbol{F_B}$ are forcing terms (assumed to be known/prescribed). In the Lagrangian frame, the spatial derivative terms in \eqref{Aeq} and \eqref{Beq} are unknown, and therefore these terms must be modeled in terms of functionals of the associated quantity, e.g $\nu\nabla^2\boldsymbol{A}$ must be modelled as some functional of $\boldsymbol{A}$. 

Concerning \eqref{Aeq}, various closure approaches have been developed, including the Recent Fluid Deformation Approximation (RFDA) \citep{chertkov1999lagrangian}, Random Gaussian Fields Closure (RGFC) \citep{wilczek2014pressure}, and the Recent Deformation of Gaussian Fields (RDGF) closure \citep{johnson2016closure}. More recently, a multi-level version of the RDGF model has emerged (referred to as ML-RDGF) that provides a model for $D_t\boldsymbol{A}$ that is valid for arbitrary Reynolds numbers. These closures for \eqref{Aeq} all lead to a model for $\boldsymbol{A}$ that can generate steady-state statistics. This is in contrast to the Restricted Euler (RE) model \citep{vieillefosse1982local,meneveau2011lagrangian} that ignores the anisotropic contribution to $\boldsymbol{H}$ and sets $\nu=0$, leading to a model for $\boldsymbol{A}$ that exhibits a finite-time singularity. Subsequent studies found that the addition of viscous effects to the RE model is not sufficient to prevent a finite-time singularity; the anisotropic pressure Hessian must also be accounted for. This could point to a potential challenge in closing \eqref{Beq}; if the closure approximation for $\kappa\nabla^2\boldsymbol{B}$ is not sufficiently accurate then the predictions from the resulting model could also generate finite-time singular solutions for $\boldsymbol{B}$ since the equation for $\boldsymbol{B}$ contains no pressure Hessian to regulate the amplification term $\boldsymbol{B\cdot A}$ that causes $\|\boldsymbol{B}\|$ to grow when $\boldsymbol{B\cdot}(\boldsymbol{B\cdot A})<0$. In this sense, developing a suitable closure for \eqref{Beq} may be more challenging than that for \eqref{Aeq}.

The RGFC model (and also the RDGF and ML-RDGF models, since they are extensions of the RGFC model) does not directly approximate the unclosed terms in \eqref{Aeq} (unlike the RFDA model) but instead closes the equation for $d_t\boldsymbol{\mathcal{A}}$, which is defined as the solution to 
\eqref{Aeq}
averaged over the subset of fluid particles that experience the same value of $\boldsymbol{A}$. The statistics of $\boldsymbol{\mathcal{A}}$ correspond to the statistics of $\boldsymbol{A}$, however, the advantage is that the terms requiring closure in the equation for $d_t\boldsymbol{\mathcal{A}}$ are conditionally averaged quantities, and therefore statistical approaches based on the properties of random Gaussian fields may be employed to close the terms in the equation. Such a procedure cannot be done when directly closing the terms in \eqref{Aeq} since these involve instantaneous quantities.

Mathematically, this procedure that replaces $D_t\boldsymbol{A}$ with $d_t\boldsymbol{\mathcal{A}}$ may be formalized as follows. Suppose that $\boldsymbol{A}$ evolves in a phase-space with time-independent coordinates $\boldsymbol{a}\in\mathbb{R}^{3\times 3}$. The PDF of $\boldsymbol{A}$ is then defined as $\mathcal{P}(\boldsymbol{a},t)\equiv \langle\delta(\boldsymbol{A}-\boldsymbol{a})\rangle$ which solves the Liouville equation
\begin{align}
\partial_t\mathcal{P}=-\boldsymbol{\nabla_a\cdot}\Big(\mathcal{P}\Big\langle D_t\boldsymbol{A}\Big\rangle_{\boldsymbol{A}=\boldsymbol{a}} \Big),
\end{align}
where $\langle\cdot\rangle_{\boldsymbol{A}=\boldsymbol{a}} $ denotes an ensemble average conditioned on $\boldsymbol{A}=\boldsymbol{a}$. We now introduce the time-dependent characteristic variable $\boldsymbol{\mathcal{A}}$ defined via 
\begin{align}
d_t\boldsymbol{\mathcal{A}}&\equiv\boldsymbol{G}(\boldsymbol{\mathcal{A}},t),\label{calAdt}\\
\boldsymbol{G}(\boldsymbol{a},t)&\equiv\Big\langle D_t\boldsymbol{A}\Big\rangle_{\boldsymbol{A}=\boldsymbol{a}}.
\end{align}
Whereas $\boldsymbol{A}$ evolves according to the instantaneous Navier-Stokes equation, $\boldsymbol{\mathcal{A}}$ evolves according to the conditionally averaged Navier-Stokes equation. Nevertheless, since $\langle D_t\boldsymbol{A}\rangle_{\boldsymbol{A}=\boldsymbol{a}}$ only involves a partial average over the flow, the field $\boldsymbol{G}$ will in general exhibit nonlinear dependence on $\boldsymbol{a}$ and $t$, and hence the trajectories $\boldsymbol{\mathcal{A}}$ generated by \eqref{calAdt} will still vary chaotically in time for a turbulent flow.

We may then define the PDF $\varrho(\boldsymbol{a},t)\equiv \langle\delta(\boldsymbol{\mathcal{A}}-\boldsymbol{a})\rangle$ which solves

\begin{align}
\partial_t\varrho&=-\boldsymbol{\nabla_a\cdot}\Big(\varrho\Big\langle d_t\boldsymbol{\mathcal{A}}\Big\rangle_{\boldsymbol{\mathcal{A}}=\boldsymbol{a}} \Big).
\end{align}
From the definition of the characteristic variable we have
\begin{align}
\Big\langle  d_t\boldsymbol{\mathcal{A}}\Big\rangle_{\boldsymbol{\mathcal{A}}=\boldsymbol{a}}=\Big\langle \boldsymbol{G}(\boldsymbol{\mathcal{A}},t)\Big\rangle_{\boldsymbol{\mathcal{A}}=\boldsymbol{a}}=\boldsymbol{G}(\boldsymbol{a},t)=\Big\langle D_t\boldsymbol{A}\Big\rangle_{\boldsymbol{A}=\boldsymbol{a}},
\end{align}
where the second equality follows since $\boldsymbol{G}(\boldsymbol{a},t)$ is not random. In view of this, if $\varrho(\boldsymbol{a},0)=\mathcal{P}(\boldsymbol{a},0)$ then it follows that $\varrho(\boldsymbol{a},t)=\mathcal{P}(\boldsymbol{a},t)\,\forall t$, and hence the statistics of $\boldsymbol{A}$ may be obtained via solutions to 
\eqref{calAdt}.

Based on the definitions above we have
\begin{align}
d_t\boldsymbol{\mathcal{A}}&=-\boldsymbol{\mathcal{A}\cdot \mathcal{A}}-\langle\boldsymbol{H}\rangle_{\boldsymbol{\mathcal{A}}}+\nu\langle\nabla^2\boldsymbol{A}\rangle_{\boldsymbol{\mathcal{A}}}+\langle\boldsymbol{F_A}\rangle_{\boldsymbol{\mathcal{A}}},\label{Acaleq}
\end{align}
where we have used the short-hand notation
\begin{align}
\langle\cdot\rangle_{\boldsymbol{\mathcal{A}}}=\langle\cdot\rangle_{\boldsymbol{A}=\boldsymbol{a}}\Big\vert_{\boldsymbol{a}=\boldsymbol{\mathcal{A}}}.
\end{align}
While closing the equation for $D_t\boldsymbol{A}$ requires approximating instantaneous quantities, closing $d_t\boldsymbol{\mathcal{A}}$ requires approximating conditionally averaged quantities. This is a major advantage since it means that powerful statistical approaches can be used to develop systematic closures for $d_t\boldsymbol{\mathcal{A}}$. This then is the approach adopted in the RGFC and RDGF models, where $\langle\boldsymbol{H}\rangle_{\boldsymbol{\mathcal{A}}}$ and $\langle\nabla^2\boldsymbol{A}\rangle_{\boldsymbol{\mathcal{A}}}$ are closed under the approximation that the velocity field $\boldsymbol{u}$ has Gaussian statistics. 

A similar approach to this may be adopted for \eqref{Beq}, but now the averaging must be conditional on both $\boldsymbol{A}=\boldsymbol{a}$ and $\boldsymbol{B}=\boldsymbol{b}$, where $(\boldsymbol{a},\boldsymbol{b})\in (\mathbb{R}^{3\times 3},\mathbb{R}^3)$ are the time-independent coordinates of the phase-space in which $\boldsymbol{A}$ and $\boldsymbol{B}$ evolve. We define
\begin{align}
d_t\boldsymbol{\mathcal{B}}&\equiv\boldsymbol{J}(\boldsymbol{\mathcal{A}},\boldsymbol{\mathcal{B}},t),\label{calBdt}\\
\boldsymbol{J}(\boldsymbol{a},\boldsymbol{b},t)&\equiv\Big\langle D_t\boldsymbol{B}\Big\rangle_{\boldsymbol{A}=\boldsymbol{a},\boldsymbol{B}=\boldsymbol{b}},
\end{align}
and then based on \eqref{Beq} this leads to
\begin{align}
d_t\boldsymbol{\mathcal{B}}&=-\boldsymbol{\mathcal{A}}^\top\boldsymbol{\cdot\mathcal{B}}+\kappa\langle\nabla^2\boldsymbol{B}\rangle_{\boldsymbol{\mathcal{A}},\boldsymbol{\mathcal{B}}}+\langle\boldsymbol{F_B}\rangle_{\boldsymbol{\mathcal{A}},\boldsymbol{\mathcal{B}}},\label{Bcaleq}\\
\langle\cdot\rangle_{\boldsymbol{\mathcal{A}},\boldsymbol{\mathcal{B}}}&=\langle\cdot\rangle_{\boldsymbol{A}=\boldsymbol{a},\boldsymbol{B}=\boldsymbol{b}}\Big\vert_{\boldsymbol{a}=\boldsymbol{\mathcal{A}},\boldsymbol{b}=\boldsymbol{\mathcal{B}}}.
\end{align}
An approach based on the RGFC or RDGF models may then be used to close the term $\langle\nabla^2\boldsymbol{B}\rangle_{\boldsymbol{\mathcal{A}},\boldsymbol{\mathcal{B}}}$ under the assumption that the scalar field $\phi$ has Gaussian statistics. Just as the solutions to 
\eqref{calAdt} can be used to construct the statistics of $\boldsymbol{A}$, so also can the solutions to
\eqref{calBdt} be used to construct the statistics of $\boldsymbol{B}$.

\subsection{\label{sec:level2_mod}Model for the velocity gradients}

In our model for $\boldsymbol{\mathcal{B}}$, the ML-RDGF model will be used to prescribe $\boldsymbol{\mathcal{A}}$, and since our model for $\boldsymbol{\mathcal{B}}$ will be based on the RDGF approach, we summarize the key ideas in this modeling approach before applying them to deriving a closed equation for $d_t\boldsymbol{\mathcal{B}}$

\subsubsection{\label{sec:level3_RDGF}RDGF closure}

The deformation tensor $\boldsymbol{D}(t,s)$ for a fluid particle with position $\boldsymbol{x}^f(t\vert\boldsymbol{X},s)$ that satisfies $\boldsymbol{x}^f(s\vert\boldsymbol{X},s)=\boldsymbol{X}$ is defined as 
\begin{align}
\boldsymbol{D}(t,s)\equiv\frac{\partial}{\partial\boldsymbol{X}}\boldsymbol{x}^f(t\vert\boldsymbol{X},s),\quad s\in[0,t],    
\end{align}
and evolves according to
\begin{align}
\partial_t\boldsymbol{D}=\boldsymbol{A\cdot D},\quad \boldsymbol{D}(0,0)=\mathbf{I},
\end{align}
where $\mathbf{I}$ is the identity matrix. While the solution to this is given by a time-ordered exponential, the RFDA approximates the solution by assuming that $\boldsymbol{A}$ is constant over a recent-deformation timescale $\tau$, with the deformation at times $s<t-\tau$ ignored. In this case, the approximate solution is 

\begin{align}
\boldsymbol{D}(t,s)&\approx \exp\Big((t-s)\boldsymbol{A} \Big),\quad t-s\in[0,\tau].\label{Dappx}
\end{align}
The idea then is as follows: the quantities $\boldsymbol{H}\equiv\boldsymbol{\nabla\nabla}p$ and $\nabla^2\boldsymbol{A}$ are evaluated along the fluid particle trajectory $\boldsymbol{x}^f(t\vert\boldsymbol{X},s)$, and they may be related to their corresponding values at the reference configuration $\boldsymbol{x}^f(s\vert\boldsymbol{X},s)=\boldsymbol{X}$ using $\boldsymbol{D}$. Using the approximate solution for $\boldsymbol{D}$ in \eqref{Dappx} and setting $s=t-\tau$ leads to
\begin{align}
\langle\boldsymbol{H}\rangle_{\boldsymbol{\mathcal{A}}}&\approx \Big\langle\boldsymbol{D}^{-\top}\boldsymbol{\cdot}\Big(\frac{\partial^2}{\partial\boldsymbol{X}\partial\boldsymbol{X}}{p}\Big)\boldsymbol{\cdot}\boldsymbol{D}^{-1}\Big\rangle_{\boldsymbol{\mathcal{A}}}=\boldsymbol{\mathcal{D}}^{-\top}\boldsymbol{\cdot}\Big\langle\frac{\partial^2}{\partial\boldsymbol{X}\partial\boldsymbol{X}}{p}\Big\rangle_{\boldsymbol{\mathcal{A}}}\boldsymbol{\cdot}\boldsymbol{\mathcal{D}}^{-1},\label{Defappx1}\\
\langle\nabla^2\boldsymbol{A}\rangle_{\boldsymbol{\mathcal{A}}}&\approx \Big\langle\boldsymbol{D}^{-\top}\boldsymbol{\cdot}\Big(\frac{\partial^2}{\partial\boldsymbol{X}\partial\boldsymbol{X}}{\boldsymbol{A}}\Big)\boldsymbol{\cdot}\boldsymbol{D}^{-1}\Big\rangle_{\boldsymbol{\mathcal{A}}}=\boldsymbol{\mathcal{D}}^{-\top}\boldsymbol{\cdot}\Big\langle\frac{\partial^2}{\partial\boldsymbol{X}\partial\boldsymbol{X}}{\boldsymbol{A}}\Big\rangle_{\boldsymbol{\mathcal{A}}}\boldsymbol{\cdot}\boldsymbol{\mathcal{D}}^{-1},\label{Defappx2}
\end{align}
where $\boldsymbol{\mathcal{D}}= \exp(\tau\boldsymbol{\mathcal{A}})$, and $(\cdot)^{-\top}$ denotes the transpose of the inverse of a tensor. The RDGF model then uses the RGFC approach \citep{wilczek2014pressure} to approximate the conditional averages in \eqref{Defappx1} and \eqref{Defappx2}.

The basic motivation for the RDGF model is that the closure approximations for $\langle\boldsymbol{H}\rangle_{\boldsymbol{\mathcal{A}}}$ and $\langle\nabla^2\boldsymbol{A}\rangle_{\boldsymbol{\mathcal{A}}}$ can be improved by applying the RGFC at $\boldsymbol{X},t-\tau$ rather than $\boldsymbol{x},t$. This is because when the RGFC is applied at $\boldsymbol{X},t-\tau$ it is then transformed under the flow map into something more realistic. For example, while the conditional averages in \eqref{Defappx1} and \eqref{Defappx2} are approximated assuming that $\boldsymbol{u}$ has Gaussian statistics, due to the Lagrangian transformation described by $\boldsymbol{\mathcal{D}}$, the resulting approximations for $\langle\boldsymbol{H}\rangle_{\boldsymbol{\mathcal{A}}}$ and $\langle\nabla^2\boldsymbol{A}\rangle_{\boldsymbol{\mathcal{A}}}$ obtained through \eqref{Defappx1} and \eqref{Defappx2} will, in general, correspond to those for a non-Gaussian field $\boldsymbol{u}$. 

The final closed equation obtained using the RDGF closure has the form \citep{johnson2016closure}
\begin{align}
d\boldsymbol{\mathcal{A}}=\boldsymbol{\mathcal{N}}_{\mathcal{A}}\{\boldsymbol{\mathcal{A}},\tau,\tau_\eta\}dt+\boldsymbol{\Sigma\cdot}d\boldsymbol{\mathcal{W}},\label{RDGF_equation}   
\end{align}
where the forcing term has been chosen to be $\langle\boldsymbol{F_A}\rangle_{\boldsymbol{\mathcal{A}}}dt=\boldsymbol{\Sigma\cdot}d\boldsymbol{\mathcal{W}}$, with $d\boldsymbol{\mathcal{W}}$  denoting a tensor-valued Wiener process, and $\boldsymbol{\Sigma}$ denoting a diffusion tensor that depends on the coefficients $D_s$ and $D_a$ which determine the growth rate of the mean-square values of the strain-rate $\boldsymbol{\mathcal{S}}\equiv (\boldsymbol{\mathcal{A}}+\boldsymbol{\mathcal{A}}^\top)/2$ and rotation-rate $\boldsymbol{\mathcal{R}}\equiv (\boldsymbol{\mathcal{A}}-\boldsymbol{\mathcal{A}}^\top)/2$ tensors. For brevity, we do not include the details of the nonlinear operator $\boldsymbol{\mathcal{N}}_{\mathcal{A}}\{\cdot \}$ which may be found in previous works \citep{johnson2016closure}.

The three unknown parameters $\tau, D_a, D_s$ are obtained by an optimization procedure that seeks those values for which the model satisfies known constraints for isotropic turbulence, namely $2\langle\|\boldsymbol{\mathcal{S}}\|^2\rangle =1/\tau_\eta^2$, and the two homogeneity relations due to \cite{betchov1956inequality}, $\langle \boldsymbol{\mathcal{A}}\boldsymbol{:\mathcal{A}}\rangle=0$, $\langle (\boldsymbol{\mathcal{A}}\boldsymbol{\cdot\mathcal{A}})\boldsymbol{:\mathcal{A}}\rangle =0$. The values obtained by this procedure are $\tau=0.1302\tau_\eta$, $D_s=0.1014/\tau_\eta^3$, and $D_a=0.0505/\tau_\eta^3$.

\subsubsection{\label{sec:level3}ML-RDGF closure}

The RDGF model described by \eqref{RDGF_equation} does not contain any dependence on $\Rey_\lambda$. To address this, a multi-level version of the RDGF closure (called ML-RDGF) was developed \citep{johnson2017turbulence}. The key idea behind this model is that in a turbulent flow, there exist velocity gradients at different scales in the flow, and the velocity gradient dynamics at different scales are coupled because of the energy cascade. Moreover, this coupling will be influenced by the fact that the energy flux through the cascade is not constant, but fluctuates in time and space. The ML-RDGF model extends the RDGF model to take this into account by replacing \eqref{RDGF_equation} with
\begin{align}
d\boldsymbol{\mathcal{A}}^{[n]}=\boldsymbol{\mathcal{N}}_{\mathcal{A}}\{\boldsymbol{\mathcal{A}}^{[n]},\tau,\tau_n\}dt -d_t(\ln\tau_n)\boldsymbol{\mathcal{A}}^{[n]}  +\boldsymbol{\Sigma}^{[n]}\boldsymbol{\cdot}d\boldsymbol{\mathcal{W}}^{[n]},\quad n=1,2,...N,\label{MLRDGF_equation}   
\end{align}
in which the $\tau_\eta$ appearing in \eqref{RDGF_equation} has been replaced by the time-dependent timescale $\tau_n(t)$, so that now $\tau=0.1302\tau_n$, $D_s=0.1014/\tau_n^3$, and $D_a=0.0505/\tau_n^3$.

Equation \eqref{MLRDGF_equation} is actually a system of $N$ coupled equations, and $\boldsymbol{\mathcal{A}}^{[n]}$ represents the velocity gradient at the $n^{th}$ level, corresponding to the velocity gradient filtered on some scale. For the first level $n=1$, the timescale is fixed $\tau_1=\beta^{N-1}\tau_\eta$, where $\beta=10$ is chosen in the model. For $n\geq 2$, $\tau_n(t)\equiv \beta^{-1}\|\boldsymbol{\mathcal{S}}^{[n-1]}\|^{-1}$ (where $\boldsymbol{\mathcal{S}}^{[n-1]}$ is the strain-rate associated with $\boldsymbol{\mathcal{A}}^{[n-1]}$), such that the time evolution of $\boldsymbol{\mathcal{A}}^{[n]}$ is coupled to the evolution at the larger scale where the velocity gradient is $\boldsymbol{\mathcal{A}}^{[n-1]}$. The solution at level $n=N$ then corresponds to the full (unfiltered) velocity gradient, i.e.~$\boldsymbol{\mathcal{A}}^{[N]}=\boldsymbol{\mathcal{A}}$. This multi-level model is capable of predicting $\boldsymbol{\mathcal{A}}$ for the discrete Reynolds numbers $\Rey_\lambda=\Rey_\lambda^{[n=1]}\beta^{N-1}$, where $\Rey_\lambda^{[n=1]}$ is the Taylor Reynolds number corresponding to the first level $n=1$, which was chosen in previous studies \citep{johnson2017turbulence} to be $\Rey_\lambda^{[n=1]}=60$. Then a modified timescale for the second level, $\tau_2$, enables the model to predict flows at arbitrary $\Rey_\lambda$.

The predictions for $\boldsymbol{\mathcal{A}}^{[N]}=\boldsymbol{\mathcal{A}}$ were shown to be in excellent agreement with DNS and experimental data \citep{johnson2017turbulence}, and revealed that the model makes robust predictions for the intermittency of $\boldsymbol{\mathcal{A}}$ up to the highest Reynolds number considered, $\Rey_\lambda= \textit{O}(10^6)$. 

\subsection{\label{sec:level2}Closure for scalar gradient equation based on RDGF}

Based on its excellent performance, the ML-RDGF model will be used to specify $\boldsymbol{\mathcal{A}}$ in the equation for the scalar gradient $\boldsymbol{\mathcal{B}}$. The RDGF closure scheme will be used to close the scalar gradient diffusion term $\langle\nabla^2\boldsymbol{B}\rangle_{\boldsymbol{\mathcal{A}},\boldsymbol{\mathcal{B}}}$. A multi-level version is not required since the effect of $\Rey_\lambda$ on the model for $\boldsymbol{\mathcal{B}}$ will already be accounted for through the use of the ML-RDGF to specify $\boldsymbol{\mathcal{A}}$ in the equation for $\boldsymbol{\mathcal{B}}$.

Analogous to \eqref{Defappx1} and \eqref{Defappx2}, under the recent fluid deformation approximation, the scalar gradient diffusion term may be expressed as
\begin{align}
\langle\nabla^2\boldsymbol{B}\rangle_{\boldsymbol{\mathcal{A}},\boldsymbol{\mathcal{B}}}\approx\boldsymbol{\mathcal{D}}^{-\top}\boldsymbol{\cdot}\Big\langle\frac{\partial^2}{\partial\boldsymbol{X}\partial\boldsymbol{X}}{\boldsymbol{B}}\Big\rangle_{\boldsymbol{\mathcal{A}},\boldsymbol{\mathcal{B}}}\boldsymbol{\cdot}\boldsymbol{\mathcal{D}}^{-1}.\label{Defappx3}
\end{align}
The Random Gaussian Fields Closure (RGFC) can be used to derive a closed expression for the conditional average appearing in this expression, leading through \eqref{Defappx3} to an RDGF closure for $\langle\nabla^2\boldsymbol{B}\rangle_{\boldsymbol{\mathcal{A}},\boldsymbol{\mathcal{B}}}$.

In the present context of the scalar field, the RGFC begins with the assumption that $\phi$ has Gaussian statistics defined in terms of the characteristic functional
\begin{align}
\Sigma^{\phi}[\lambda(\boldsymbol{x})]=\exp\Bigg[-\frac{1}{2}\int_{\mathbb{R}^3}\int_{\mathbb{R}^3}\lambda(\boldsymbol{x})R^{\phi}(\boldsymbol{x},\boldsymbol{x}')\lambda(\boldsymbol{x}')\,d\boldsymbol{x}\, d\boldsymbol{x}' \Bigg],\label{phi_char}
\end{align}
(time label is suppressed here and in what follows for simplicity) where $\lambda(\boldsymbol{x})$ is the Fourier variable conjugate to $\phi(\boldsymbol{x})$, and $R^{\phi}(\boldsymbol{x},\boldsymbol{x}')\equiv\langle \phi(\boldsymbol{x})\phi(\boldsymbol{x}')\rangle$, which for isotropic turbulence has the form
\begin{align}
R^{\phi}(\boldsymbol{x},\boldsymbol{x}')=R^{\phi}(r)=\langle\phi^2\rangle f_\phi(r),\quad \boldsymbol{r}\equiv \boldsymbol{x}-\boldsymbol{x}',\quad r\equiv\|\boldsymbol{r}\|,\label{iso_Rphi}
\end{align}
where $f_\phi(r)$ is the scalar spatial correlation function. Following the work of \cite{wilczek2014pressure}, the characteristic function for the scalar gradient $\boldsymbol{B}\equiv\boldsymbol{\nabla}\phi$ is obtained from \eqref{phi_char} as
\begin{align}
\Sigma^{\boldsymbol{B}}[\boldsymbol{\mu}(\boldsymbol{x})]=\exp\Bigg[-\frac{1}{2}\int_{\mathbb{R}^3}\int_{\mathbb{R}^3}\boldsymbol{\mu}(\boldsymbol{x})\boldsymbol{\cdot}\boldsymbol{R}^{\boldsymbol{B}}(\boldsymbol{x},\boldsymbol{x}')\boldsymbol{\cdot}\boldsymbol{\mu}(\boldsymbol{x}')\,d\boldsymbol{x}\, d\boldsymbol{x}' \Bigg],\label{B_char}
\end{align}
where $\boldsymbol{\mu}$ is the Fourier variable conjugate to $\boldsymbol{B}$, and $\boldsymbol{R}^{\boldsymbol{B}}(\boldsymbol{x},\boldsymbol{x}')\equiv\langle\boldsymbol{B}(\boldsymbol{x})\boldsymbol{B}'(\boldsymbol{x}')\rangle$, which is related to $R^{\phi}(\boldsymbol{x},\boldsymbol{x}')$ through
\begin{align}
\boldsymbol{R}^{\boldsymbol{B}}(\boldsymbol{x},\boldsymbol{x}')=\langle\boldsymbol{\nabla} \phi(\boldsymbol{x})\boldsymbol{\nabla}'\phi(\boldsymbol{x}')\rangle  = \boldsymbol{\nabla}\boldsymbol{\nabla}'R^{\phi}(\boldsymbol{x},\boldsymbol{x}').
\end{align}
For an isotropic scalar field where \eqref{iso_Rphi} applies, we then have
\begin{align}
\boldsymbol{R}^{\boldsymbol{B}}(\boldsymbol{x},\boldsymbol{x}')=\boldsymbol{R}^{\boldsymbol{B}}(\boldsymbol{r})=\langle\phi^2\rangle\Bigg[\Bigg(\frac{f_\phi'(r)}{r}-f_\phi^{''}(r)\Bigg)\frac{\boldsymbol{rr}}{r^2}-\frac{f_\phi'(r)}{r}\mathbf{I} \Bigg],\label{Bcov}
\end{align}
where prime superscripts denote differentiation with respect to $r$.

According to \eqref{B_char}, the statistics of $\boldsymbol{B}$ are described by a Gaussian characteristic functional when $\phi$ is assumed to be Gaussian. Just as in previous works \citep{wilczek2014pressure,johnson2016closure} when deriving a closure model for $D_t\boldsymbol{\mathcal{A}}$, it is acknowledged that the statistics of $\boldsymbol{B}$ are not Gaussian in a real turbulent flow (nor are they Gaussian even when $\boldsymbol{u}$ is Gaussian \citep{falkovich01}). The motivation for this choice is simply that it is the only option when deriving a statistical field closure. It must be appreciated, however, that only with respect to the closure of $\langle\nabla^2\boldsymbol{B}\rangle_{\boldsymbol{\mathcal{A}},\boldsymbol{\mathcal{B}}}$, are the statistics of $\boldsymbol{B}$ approximated as being Gaussian; the model that follows from this choice nevertheless generates statistics for $\boldsymbol{\mathcal{B}}$ that are highly non-Gaussian (as will be shown later).

The specific term to be closed using RGFC is
\begin{align}
\Big\langle\frac{\partial^2}{\partial\boldsymbol{X}\partial\boldsymbol{X}}{\boldsymbol{B}}\Big\rangle_{\boldsymbol{a},\boldsymbol{b}}=\Big\langle\frac{\partial^2}{\partial\boldsymbol{X}\partial\boldsymbol{X}}{\boldsymbol{B}}(\boldsymbol{x}^f(t\vert\boldsymbol{X},s),t)\Big\rangle_{{\boldsymbol{A}}(\boldsymbol{x}^f(t\vert\boldsymbol{X},s),t)=\boldsymbol{a},{\boldsymbol{B}}(\boldsymbol{x}^f(t\vert\boldsymbol{X},s),t)=\boldsymbol{b}},
\end{align}
which is evaluated at $\boldsymbol{a}=\boldsymbol{\mathcal{A}}, \boldsymbol{b}=\boldsymbol{\mathcal{B}}$ in \eqref{Defappx3}. Just as the RFDA assumes ${\boldsymbol{A}}(\boldsymbol{x}^f(t\vert\boldsymbol{X},s),t)\approx {\boldsymbol{A}}(\boldsymbol{X},s)$ for $t-s\in[0,\tau]$, we also assume ${\boldsymbol{B}}(\boldsymbol{x}^f(t\vert\boldsymbol{X},s),t)\approx {\boldsymbol{B}}(\boldsymbol{X},s)$ for $t-s\in[0,\tau]$, and inserting this yields 
\begin{align}
\Big\langle\frac{\partial^2}{\partial\boldsymbol{X}\partial\boldsymbol{X}}{\boldsymbol{B}}\Big\rangle_{\boldsymbol{a},\boldsymbol{b}}\approx\Big\langle\frac{\partial^2}{\partial\boldsymbol{X}\partial\boldsymbol{X}}{\boldsymbol{B}}(\boldsymbol{X},s)\Big\rangle_{{\boldsymbol{A}}(\boldsymbol{X},s)=\boldsymbol{a},{\boldsymbol{B}}(\boldsymbol{X},s)=\boldsymbol{b}},
\end{align}
which puts the conditional average into a form to which the RGFC procedure \citep{wilczek2014pressure} can be applied.

Before proceeding, we note that although the argument of the conditional average\[\Big\langle\frac{\partial^2}{\partial\boldsymbol{X}\partial\boldsymbol{X}}{\boldsymbol{B}}(\boldsymbol{X},s)\Big\rangle_{{\boldsymbol{A}}(\boldsymbol{X},s)=\boldsymbol{a},{\boldsymbol{B}}(\boldsymbol{X},s)=\boldsymbol{b}}\]does not contain $\boldsymbol{A}$ but only $\boldsymbol{B}$, the conditionality on $\boldsymbol{A}(\boldsymbol{X},s)=\boldsymbol{a}$ cannot formally be removed. This is because since the evolution of $\boldsymbol{B}$ depends upon $\boldsymbol{A}$ (see \eqref{Beq}), then the value of \[\frac{\partial^2}{\partial\boldsymbol{X}\partial\boldsymbol{X}}{\boldsymbol{B}}\]will not be uncorrelated from $\boldsymbol{A}$, in general. While a closure for the full conditional average can be obtained using the RDGF approach, in order to simplify the closure analysis the following approximation is made
\begin{align}
\Big\langle\frac{\partial^2}{\partial\boldsymbol{X}\partial\boldsymbol{X}}{\boldsymbol{B}}(\boldsymbol{X},s)\Big\rangle_{{\boldsymbol{A}}(\boldsymbol{X},s)=\boldsymbol{a},{\boldsymbol{B}}(\boldsymbol{X},s)=\boldsymbol{b}}\approx \Big\langle\frac{\partial^2}{\partial\boldsymbol{X}\partial\boldsymbol{X}}{\boldsymbol{B}}(\boldsymbol{X},s)\Big\rangle_{{\boldsymbol{B}}(\boldsymbol{X},s)=\boldsymbol{b}}.\label{to_close}
\end{align}
It is crucial to note, however, that the overall closure for $\langle\nabla^2\boldsymbol{B}\rangle_{\boldsymbol{\mathcal{A}},\boldsymbol{\mathcal{B}}}$ does partially include the effect of the conditioning upon $\boldsymbol{\mathcal{A}}$, since the right hand side of \eqref{Defappx3} contains $\boldsymbol{\mathcal{D}}$ and not $\boldsymbol{D}$ precisely because of the conditionality in the averaging operator on the left hand side of \eqref{Defappx3}. Therefore, the closure for $\langle\nabla^2\boldsymbol{B}\rangle_{\boldsymbol{\mathcal{A}},\boldsymbol{\mathcal{B}}}$ captures some of the dependency of the scalar diffusion on the local velocity gradients in the flow.

With the statistics of $\boldsymbol{B}$ prescribed using \eqref{B_char} and the covariance tensor for $\boldsymbol{B}$ prescribed using \eqref{Bcov}, a closure for \eqref{to_close}, and hence $\langle\nabla^2\boldsymbol{B}\rangle_{\boldsymbol{\mathcal{A}},\boldsymbol{\mathcal{B}}}$, may be obtained following the same steps as in the work of \cite{wilczek2014pressure}. The basic steps are as follows. First, using the approach described in Appendix A of the work of \cite{wilczek2014pressure}, the unclosed quantity is re-written in terms of a two-point quantity 
\begin{align}
\Big\langle\frac{\partial^2}{\partial\boldsymbol{X}\partial\boldsymbol{X}}{\boldsymbol{B}}(\boldsymbol{X},s)\Big\rangle_{{\boldsymbol{B}}(\boldsymbol{X},s)=\boldsymbol{b}}=\lim_{r\to 0}\frac{\partial^2}{\partial\boldsymbol{r}\partial\boldsymbol{r}}\Big\langle \boldsymbol{B}(\boldsymbol{X}+\boldsymbol{r},s)\Big\rangle_{{\boldsymbol{B}}(\boldsymbol{X},s)=\boldsymbol{b}}.
\end{align}
Applying the steps outlined in Appendix B of \cite{wilczek2014pressure} to the scalar field we obtain
\begin{align}
\Big\langle \boldsymbol{B}(\boldsymbol{X}+\boldsymbol{r},s)\Big\rangle_{{\boldsymbol{B}}(\boldsymbol{X},s)=\boldsymbol{b}}=\boldsymbol{R}^{\boldsymbol{B}}(\boldsymbol{r})\boldsymbol{\cdot}[\boldsymbol{R}^{\boldsymbol{B}}(\boldsymbol{0})]^{-1}\boldsymbol{\cdot b},
\end{align}
where
\begin{align}
[\boldsymbol{R}^{\boldsymbol{B}}(\boldsymbol{0})]^{-1}=\frac{\mathbf{I}}{\langle\phi^2\rangle f_\phi^{''}(0)}.
\end{align}
The resulting expression
\begin{align}
\Big\langle\frac{\partial^2}{\partial\boldsymbol{X}\partial\boldsymbol{X}}{\boldsymbol{B}}(\boldsymbol{X},s)\Big\rangle_{{\boldsymbol{B}}(\boldsymbol{X},s)=\boldsymbol{b}}=\lim_{r\to 0}\Bigg(\frac{\partial^2}{\partial\boldsymbol{r}\partial\boldsymbol{r}}\boldsymbol{R}^{\boldsymbol{B}}(\boldsymbol{r})\Bigg)\boldsymbol{\cdot}[\boldsymbol{R}^{\boldsymbol{B}}(\boldsymbol{0})]^{-1}\boldsymbol{\cdot b},
\end{align}
may then be computed using \eqref{Bcov}, and when the result is evaluated at $\boldsymbol{b}=\boldsymbol{\mathcal{B}}$ and substituted into \eqref{Defappx3}, the closed expression obtained is
\begin{align}
 \kappa\langle\nabla^2\boldsymbol{B}\rangle_{\boldsymbol{\mathcal{A}},\boldsymbol{\mathcal{B}}}&\approx \delta_\mathcal{B}\Big(\boldsymbol{\mathcal{C}}_R^{-1}\boldsymbol{\cdot\mathcal{B}}+\boldsymbol{\mathcal{C}}_R^{-\top}\boldsymbol{\cdot\mathcal{B}} +\mathrm{tr}(\boldsymbol{\mathcal{C}}_R^{-1})\boldsymbol{\mathcal{B}}\Big),\label{closed_B_term}\\
 \delta_\mathcal{B}&\equiv\frac{\kappa}{3}\frac{f_\phi^{''''}(0)}{f_\phi^{''}(0)},
\end{align}
where $\boldsymbol{\mathcal{C}}_R^{-1}\equiv\boldsymbol{\mathcal{D}}^{-1}\boldsymbol{\cdot}\boldsymbol{\mathcal{D}}^{-\top}$ is the inverse of the right Cauchy-Green tensor.

In the work of \cite{johnson2016closure}, the coefficient analogous to $\delta_\mathcal{B}$ in the closure for $\langle\nabla^2\boldsymbol{A}\rangle_{\boldsymbol{\mathcal{A}}}$, namely $\delta_\mathcal{A}$, was estimated based on the enstrophy production-dissipation balance at steady-state. The same procedure can be applied to approximate $\delta_\mathcal{B}$ based on the steady-state production-dissipation balance $\langle\boldsymbol{S:BB}\rangle=-\kappa\langle\|\boldsymbol{\nabla B}\|^2 \rangle $, leading to
\begin{align}
\delta_\mathcal{B}&\approx \frac{1}{9\tau_\eta}\mathrm{tr}(\boldsymbol{\mathcal{C}}_L)\gamma_\mathcal{B},\label{deltaBappx}\\
\gamma_\mathcal{B}&\equiv \tau_\eta\frac{\langle\boldsymbol{S:BB}\rangle}{\langle \|\boldsymbol{B}\|^2\rangle},
\end{align}
where $\boldsymbol{\mathcal{C}}_L\equiv\boldsymbol{\mathcal{D}}\boldsymbol{\cdot}\boldsymbol{\mathcal{D}}^\top$ is the left Cauchy-Green tensor. 

Whereas the RDGF model for $\langle\nabla^2\boldsymbol{A}\rangle_{\boldsymbol{\mathcal{A}}}$ is nonlinear in $\boldsymbol{\mathcal{A}}$ due to the contributions from $\boldsymbol{\mathcal{C}}_L$ and $\boldsymbol{\mathcal{C}}_R$, the closure for $ \langle\nabla^2\boldsymbol{B}\rangle_{\boldsymbol{\mathcal{A}},\boldsymbol{\mathcal{B}}}$ in \eqref{closed_B_term} is linear in $\boldsymbol{\mathcal{B}}$. This could potentially suggest an issue with \eqref{closed_B_term} since it is known that models for $\langle\nabla^2\boldsymbol{A}\rangle_{\boldsymbol{\mathcal{A}}}$ that are linear in $\boldsymbol{\mathcal{A}}$ can suffer from finite-time singularities, depending on the initial conditions \citep{martin1998dynamics}. However, given that the production term in the equation for $\boldsymbol{\mathcal{B}}$ is $-\boldsymbol{\mathcal{A}}^\top\boldsymbol{\cdot\mathcal{B}}$, then the dependence of the closure in \eqref{closed_B_term} on $\boldsymbol{\mathcal{C}}_L$ and $\boldsymbol{\mathcal{C}}_R$ may indirectly prevent singular growth of $\|\boldsymbol{\mathcal{B}}\|$, at least in regions where this growth is associated with large values of $\|\boldsymbol{\mathcal{A}}\|$, because in those regions the growth of $\boldsymbol{\mathcal{B}}$ could be modulated through $\boldsymbol{\mathcal{C}}_L$ and $\boldsymbol{\mathcal{C}}_R$. Moreover, we performed tests where in the closure for $ \langle\nabla^2\boldsymbol{B}\rangle_{\boldsymbol{\mathcal{A}},\boldsymbol{\mathcal{B}}}$, we set $\boldsymbol{\mathcal{D}}=\mathbf{I}$, i.e.~so that the recent deformation mapping was removed. Simulations of the model using this blew up, showing that the contributions in \eqref{closed_B_term} involving $\boldsymbol{\mathcal{D}}$ do indeed play a key indirect role in preventing singular growth of $ \boldsymbol{\mathcal{B}}$.

In view of \eqref{deltaBappx}, any dependence of the closed expression for $\kappa\langle\nabla^2\boldsymbol{B}\rangle_{\boldsymbol{\mathcal{A}},\boldsymbol{\mathcal{B}}}$ on $\kappa$ is contained within $\gamma_\mathcal{B}$, and this term must be specified using data. However, in order for the model to depend upon $\kappa$ and hence $Sc$, then the data must specify $\gamma_\mathcal{B}$ as a function of $Sc$, which is not desirable. An alternative approach is as follows: The closure approximation in \eqref{closed_B_term} is linear in $\boldsymbol{\mathcal{B}}$, and $1/\delta_\mathcal{B}$ may be regarded as a linear relaxation timescale for $\boldsymbol{\mathcal{B}}$, describing how the diffusion term $\kappa\langle\nabla^2\boldsymbol{B}\rangle_{\boldsymbol{\mathcal{A}},\boldsymbol{\mathcal{B}}}$ causes $\boldsymbol{\mathcal{B}}$ to relax to its equilibrium value. In view of this, $1/\delta_\mathcal{B}$ should scale with the small-scale scalar timescale $\tau_\phi$, and for $Sc\geq 1$ this timescale can be estimated using the Batchelor scale \citep{batchelor1959small,donzis2005scalar} as $\tau_\phi\sim \tau_\eta Sc^{-1/3}$, while for $Sc<1$ the Corrsin scale \citep{corrsin1951spectrum} may be used to obtain $\tau_\phi\sim \tau_\eta Sc^{-1/2}$ . In view of this, the $Sc$ dependence may be explicitly accounted for in $\delta_\mathcal{B}$ by using
\begin{align}
\delta_\mathcal{B}&\approx \frac{Sc^{-\xi}}{9\tau_\eta}\mathrm{tr}(\boldsymbol{\mathcal{C}}_L)\alpha_\mathcal{B},\label{deltaBappx2}\\
\alpha_\mathcal{B}&\equiv \tau_\eta\frac{\langle\boldsymbol{S:BB}\rangle}{\langle \|\boldsymbol{B}\|^2\rangle}\Bigg\vert_{Sc=1},\label{alphaB}
\end{align}
where $\xi=1/2$ for $Sc<1$, and $\xi=1/3$ for $Sc\geq 1$. From our DNS (see \S\ref{NSims}) we obtain the value $\alpha_\mathcal{B}\approx -0.32$. However, in anticipation of results to be shown later, we note that using this fixed value for $\alpha_B$ yields a model whose predictions are not accurate (see figure \ref{PDF_B2}). Therefore, a modification will be introduced wherein $\alpha_\mathcal{B}$ is specified based on the local scalar gradient production along the trajectory history of the particle, as in equation \eqref{alphaBnew}, which dramatically improves the predictions from the model (cf.~figure \ref{PDF_B2_new}).

Analogous to the equation for $\boldsymbol{\mathcal{A}}$, the forcing term in the scalar gradient equation is chosen to be $\langle\boldsymbol{F_B}\rangle_{\boldsymbol{\mathcal{A}},\boldsymbol{\mathcal{B}}}dt=\sigma d\boldsymbol{\mathcal{W}}$, where now $d\boldsymbol{\mathcal{W}}$ is a vector-valued Wiener process with increments defined by
\begin{align}
\langle d\boldsymbol{\mathcal{W}}\rangle&=\boldsymbol{0},\\
\langle d\boldsymbol{\mathcal{W}}d\boldsymbol{\mathcal{W}}\rangle&=\mathbf{I}dt.
\end{align}
Since the equation for $\boldsymbol{\mathcal{B}}$ is linear, the statistics scale with the forcing amplitude $\sigma$, and therefore $\sigma$ may be chosen arbitrarily if the results generated by the model are suitably normalized. Using this forcing term, the final form of the model equation is written as a stochastic differential equation
\begin{align}
d\boldsymbol{\mathcal{B}}&\approx-\boldsymbol{\mathcal{A}}^\top\boldsymbol{\cdot\mathcal{B}}dt+\delta_\mathcal{B}\Big(\boldsymbol{\mathcal{C}}_R^{-1}\boldsymbol{\cdot\mathcal{B}}+\boldsymbol{\mathcal{C}}_R^{-\top}\boldsymbol{\cdot\mathcal{B}} +\mathrm{tr}(\boldsymbol{\mathcal{C}}_R^{-1})\boldsymbol{\mathcal{B}}\Big)dt+\sigma d\boldsymbol{\mathcal{W}}.\label{Beq_closed_forcing}
\end{align}
For applications where there is a mean scalar gradient, the forcing term could be replaced by the term describing the production of fluctuating scalar gradients due to the imposed mean scalar gradient.

Finally, just as the ML-RDGF model replaces the constant timescale for $\boldsymbol{\mathcal{A}}^{[N]}$ with the fluctuating timescale $\tau_N(t)$, the Kolmogorov timescale $\tau_\eta$ that appears in \eqref{deltaBappx2} should also for consistency be replaced by $\tau_N(t)$ (since the ML-RDGF model is being used to specify $\boldsymbol{\mathcal{A}}$ in \eqref{Beq_closed_forcing}). In other words \eqref{deltaBappx2} is to be replaced by
\begin{align}
\delta_\mathcal{B}&\approx \frac{Sc^{-\xi}}{9\tau_N(t)}\mathrm{tr}(\boldsymbol{\mathcal{C}}_L)\alpha_\mathcal{B}.\label{deltaBappx3}
\end{align}
This ensures that for $Sc=1$, the local timescale on which both $\boldsymbol{\mathcal{B}}$ and $\boldsymbol{\mathcal{A}}$ fluctuate is $ O(\tau_N(t))$.

\section{Numerical Simulations}\label{NSims}


We test the model predictions against data from Direct Numerical Simulation (DNS) of a passive scalar field advected by an incompressible, three-dimensional, statistically steady and isotropic turbulent velocity field.
The DNS code uses a standard Fourier pseudo-spectral method \citep{canuto1988spectral} to solve the discretized Navier-Stokes and passive scalar equations on a triply-periodic cubic domain. The required Fourier transforms are executed in parallel using the P3DFFT library \citep{pekurovsky2012p3dfft} and the aliasing error is removed via the $3/2$ rule \citep{canuto1988spectral}.
The code is described in further detail in \cite{Carbone2019}.

The non-dimensional governing equation for the passive scalar in Fourier space reads
\begin{align}
\partial_t \hat{\phi} + \textrm{i} \boldsymbol{k}\bcdot\widehat{\boldsymbol{u}\phi} = 
-\kappa\|\boldsymbol{k}\|^2\hat{\phi}
+ \hat{F},
\label{eq_scal_field}
\end{align}
where the hat indicates a Fourier transform, ``$\textrm{i}$'' is the imaginary unit, $\boldsymbol{u}(\boldsymbol{x},t)$ is the turbulent velocity field, $\phi(\boldsymbol{x},t)$ is the passive scalar field and $\hat{\phi}(\boldsymbol{k},t)$ denotes its spatial Fourier transform.
For consistency with the stochastic model, the passive scalar field is driven by a stochastic forcing, such that the complex forcing $\hat{F}$ is Gaussian and white in time, with correlation
\begin{align}
\left\langle \hat{F}(\boldsymbol{k},t)\hat{F}(\boldsymbol{k}',t')  \right\rangle = 2\sigma_0^2 \|\boldsymbol{k}\|^2 \delta(\boldsymbol{k}+\boldsymbol{k}')\delta(t-t'), \; 0<\|\boldsymbol{k}\|<k_f.
\label{eq_forcing_correlations}
\end{align}
The forcing is confined to the wavevectors $\boldsymbol{k}$ within a sphere of radius $k_f$, and we choose $k_f=\sqrt{7}$ since it yields large-scale statistics that are close to being isotropic.
Finally, the spectrum of the forcing $|\hat{F}|^2(k)$ scales as $\|\boldsymbol{k}\|^2$, compatible with energy equipartition among the smallest Fourier modes.
The constant parameter $\sigma_0$ in \eqref{eq_forcing_correlations} regulates the dissipation rate and it is adjusted to yield an appropriate Kolmogorov scale $\eta$ prescribed by the spatial resolution requirements.
We simulate a flow with $\Rey_\lambda=100$ and $Sc=1$. The spatial resolution is $k_{\max}\eta\simeq 3$, with $k_{\max}=N/2$ being the maximum resolved wavenumber, for all three simulations.
The time integration of equation \eqref{eq_scal_field} is performed by means of a second-order Runge-Kutta scheme designed for stochastic differential equations \citep{Honeycutt1992} and the CFL number stays below 0.3.


Regarding the numerical  simulations of the model, we solve the scalar gradient equation \eqref{Beq_closed_forcing} through a second-order predictor-corrector method \citep{kloeden2018numerical} with a time step $dt=0.05\tau_{\eta}$. Each level of the ML-RDGF model equation was solved using its own appropriate time step, namely, level $n$ was solved using $dt=0.05\langle \tau_n(t)\rangle$. Tests were performed using smaller time steps and these tests indicated that the aforementioned time steps were small enough to achieve convergence of the results.

We numerically solve the model equation \eqref{Beq_closed_forcing} for $\Rey_\lambda=100, 300, 500$ and the results from these will be referred to as M1, M2, and M3 in the results section. While M1 is designed to match the DNS, M2 and M3 will be used to explore the model's ability to capture the effect of $\Rey_\lambda$ on the scalar gradient statistics.

\section{Results and Discussion}

\subsection{\label{sec:level2_A} Velocity Gradients}

In our scalar gradient model, the velocity gradients are specified using the ML-RDGF model\citep{johnson2017turbulence}. We therefore begin by comparing the predictions of this model against the DNS data in order to assess its accuracy in predicting the statistics $\boldsymbol{\mathcal{A}}$, since any inaccuracies in this model will in turn lead to inaccuracies in our model for the scalar gradients.

{\vspace{0mm}\begin{figure}
		\centering
		\subfloat[]
		{\begin{overpic}
				[trim = 5mm 60mm 0mm 60mm,scale=0.3,clip,tics=20]{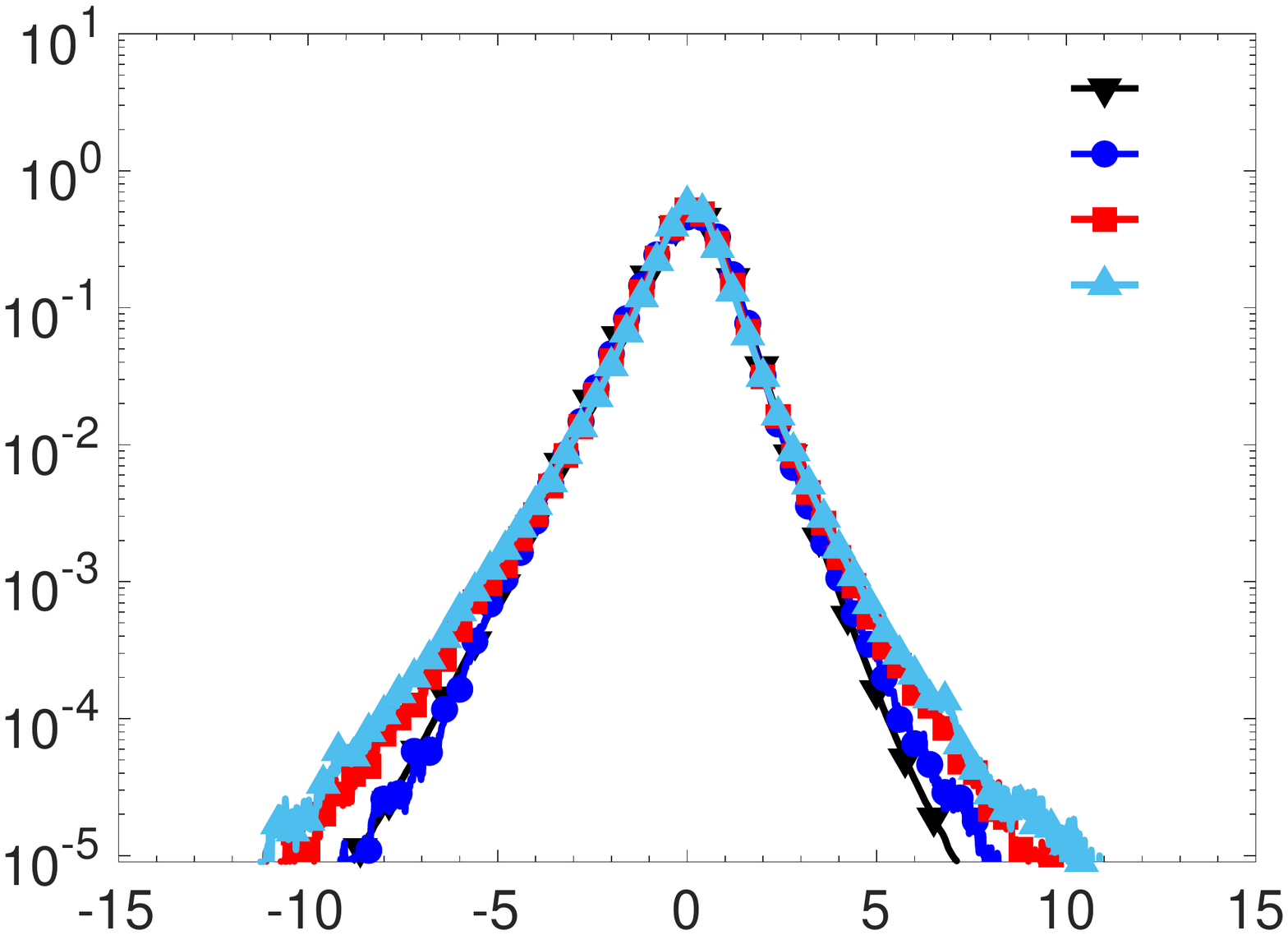}
				\put(70,0){$a_{11}/a_{11,rms}$}
				\put(-13,60){\rotatebox{90}{PDF}}
				\put(122,115){DNS}
				\put(122,107){M1}
				\put(122,98){M2}
			    \put(122,89){M3}
		\end{overpic}}
		\centering
		\subfloat[]
		{\hspace{2mm}\begin{overpic}
				[trim = 5mm 60mm 0mm 60mm,scale=0.3,clip,tics=20]{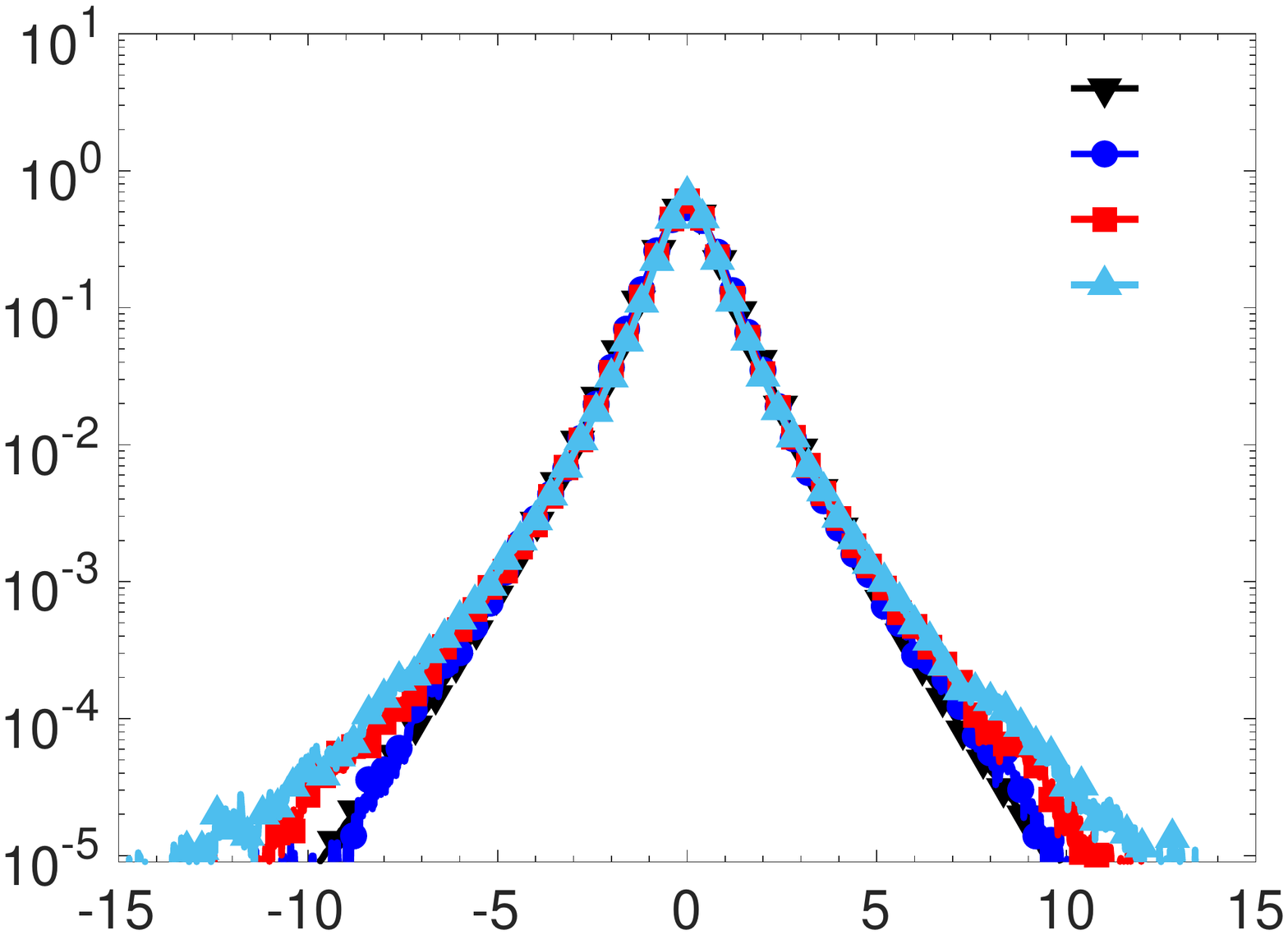}
				\put(70,0){$a_{12}/a_{12,rms}$}
				\put(-13,60){\rotatebox{90}{PDF}}
				\put(122,115){DNS}
				\put(122,107){M1}
				\put(122,98){M2}
			    \put(122,89){M3}				
		\end{overpic}}		
		\caption{PDFs of (a) longitudinal $a_{11}/a_{11,rms}$ and (b) transverse $a_{12}/a_{12,rms}$ velocity gradients.} 
		\label{PDF_aij}
\end{figure}}

In Figure \ref{PDF_aij} we compare the PDFs of the longitudinal and transverse components of the velocity gradient predicted by the ML-RDGF model with the DNS results. (Recall that $\boldsymbol{a}$ and $\boldsymbol{b}$ are the phase-space variables conjugate to $\boldsymbol{\mathcal{A}}$ and $\boldsymbol{\mathcal{B}}$, respectively, and we use the subscripts ``$rms$'' and ``$av$'' to denote the root-mean-square and averaged values of the variable under consideration). One can see that the PDFs for the longitudinal gradients (Figure \ref{PDF_aij}(a)) are negatively skewed while the PDFs for the transverse gradients are symmetric (Figure \ref{PDF_aij}(b)). The former is associated with the self-amplification of the velocity gradients \citep{tsinober}, while the latter is a constraint due to isotropy. The predictions from M1 are in excellent agreement with the DNS data, and the model makes realistic predictions at the higher $\Rey_\lambda$ to which M2 and M3 correspond. The ability of the ML-RDGF model to accurately predict the components of the velocity gradient, and realistic intermittency trends with increasing $\Rey_\lambda$ (and over a much larger range of $\Rey_\lambda$ than considered here) were previously demonstrated in detail in the original ML-RDGF paper of \cite{johnson2017turbulence}.

Further insight into the ability of the ML-RDGF model to predict the velocity gradient dynamics can be obtained by considering its predictions for the velocity gradient invariants $Q\equiv-\boldsymbol{a}\boldsymbol{:a}/2$ and $R\equiv-(\boldsymbol{a \cdot} \boldsymbol{a}) \boldsymbol{:a}/3$. The invariant $Q$ measures the relative strength of the local strain rate and vorticity in the flow, with $Q>0$ denoting vorticity-dominated regions of the flow, while $R$ measures the relative importance of the local strain self-amplification and enstrophy production, with $R<0$ denoting regions dominated by enstrophy production \citep{tsinober}. The joint PDFs of $Q$ and $R$ from the ML-RDGF model for M1, M2, M3 are presented in Figure \ref{PDF_QR_A}, along with the DNS results for comparison. As observed previously \citep{johnson2017turbulence}, the ML-RDGF captures the main features of the $Q,R$ joint-PDF well, including the signature sheared-drop shape of the joint-PDF, which is preserved by the model as $\Rey_\lambda$ is increased. The joint-PDF contours extend along the right Viellefosse tail, and the PDF is concentrated in the quadrants $Q>0, R<0$ and $Q<0, R>0$, consistent with previous studies \citep{johnson2016closure, johnson2017turbulence}. For M1, the joint-PDF is more compact compared to the DNS, indicating that the model underpredicts the probability of large values of $Q,R$. The agreement between the ML-RDGF model and DNS data was shown to be better in \cite{johnson2017turbulence}, however, their DNS data was for $\Rey_\lambda=430$, indicating that the quantitive accuracy of the model is better at higher $\Rey_\lambda$. Considering the results for M2 and M3 shows that as $\Rey_\lambda$ increases, the contours of the joint-PDF spread out in the $Q,R$ phase-space, showing that the model captures the effect of $\Rey_\lambda$ on the intermittency in the flow. 

{\vspace{0mm}\begin{figure}
		\centering
		\subfloat[]
		{\begin{overpic}
				[trim = 0mm 60mm 10mm 60mm,scale=0.3,clip,tics=20]{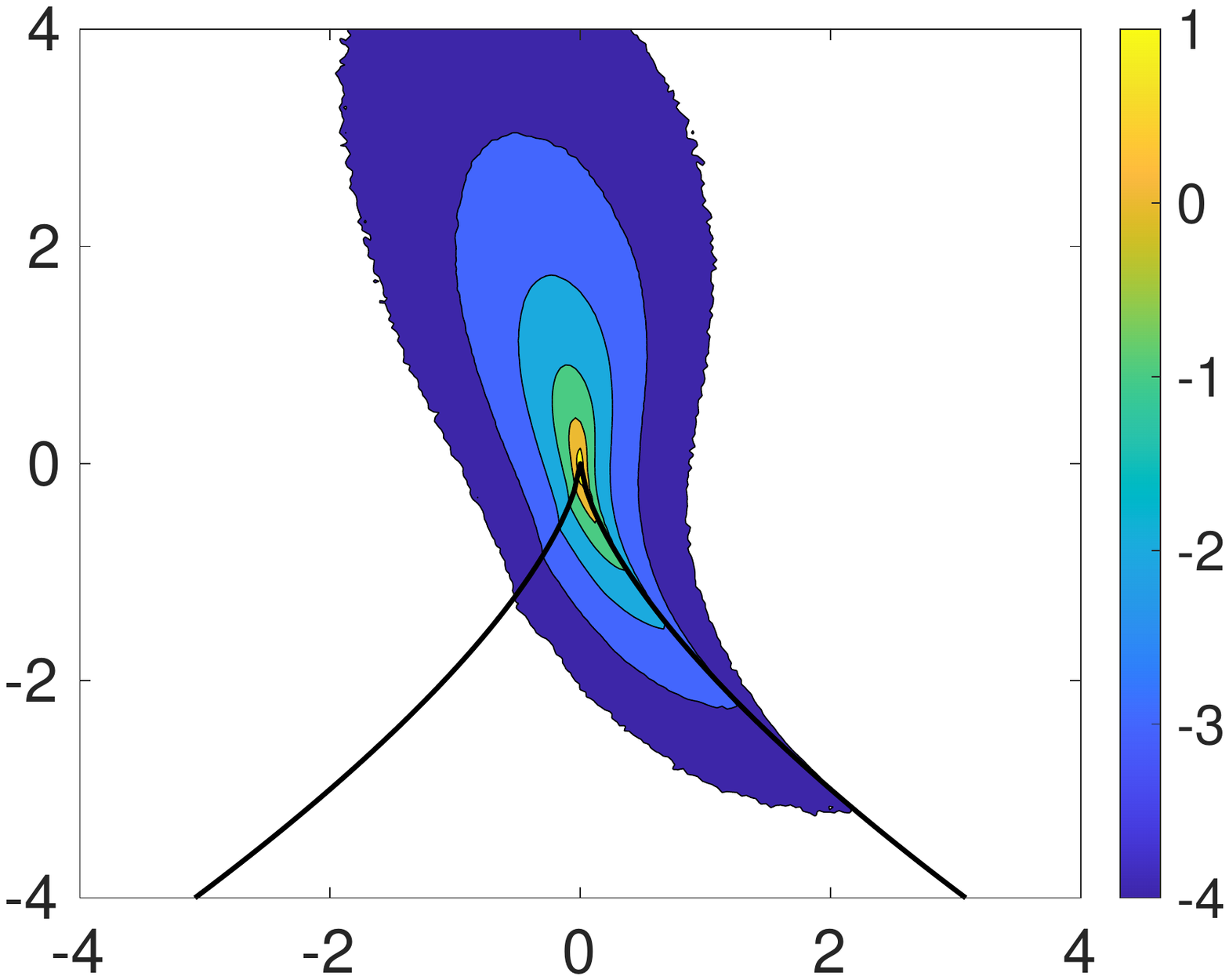}
				\put(75,-1){$R/\tilde{Q}_{av}^{3/2}$}
				\put(8,55){\rotatebox{90}{$Q/\tilde{Q}_{av}$}}				
		\end{overpic}}
		\centering
		\subfloat[]
		{\begin{overpic}
				[trim =0mm 60mm 10mm 60mm,scale=0.3,clip,tics=20]{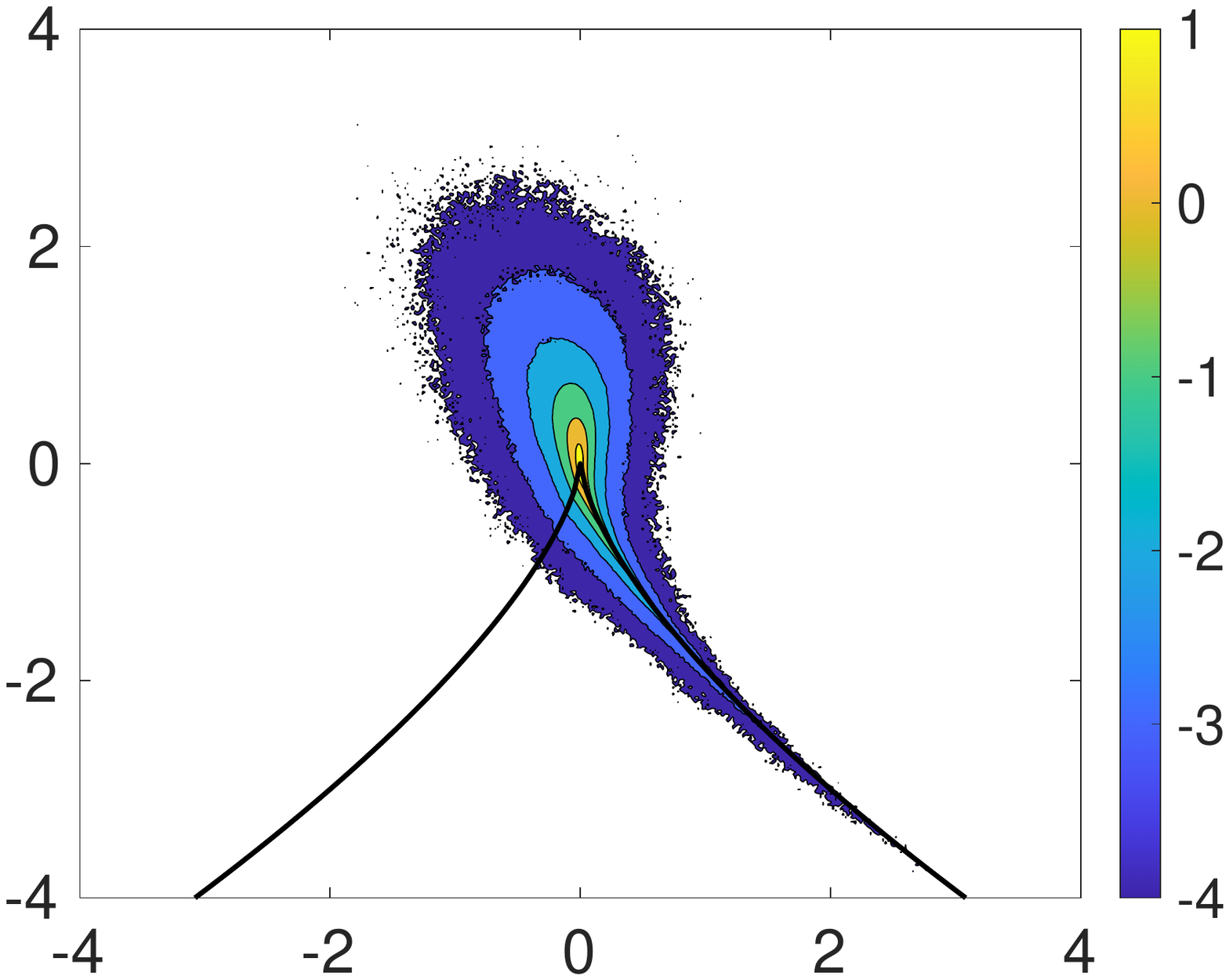}
				\put(75,-1){$R/\tilde{Q}_{av}^{3/2}$}
				\put(8,55){\rotatebox{90}{$Q/\tilde{Q}_{av}$}}								
		\end{overpic}}\\
		\centering
		\subfloat[]
		{\begin{overpic}
				[trim = 0mm 60mm 10mm 60mm,scale=0.3,clip,tics=20]{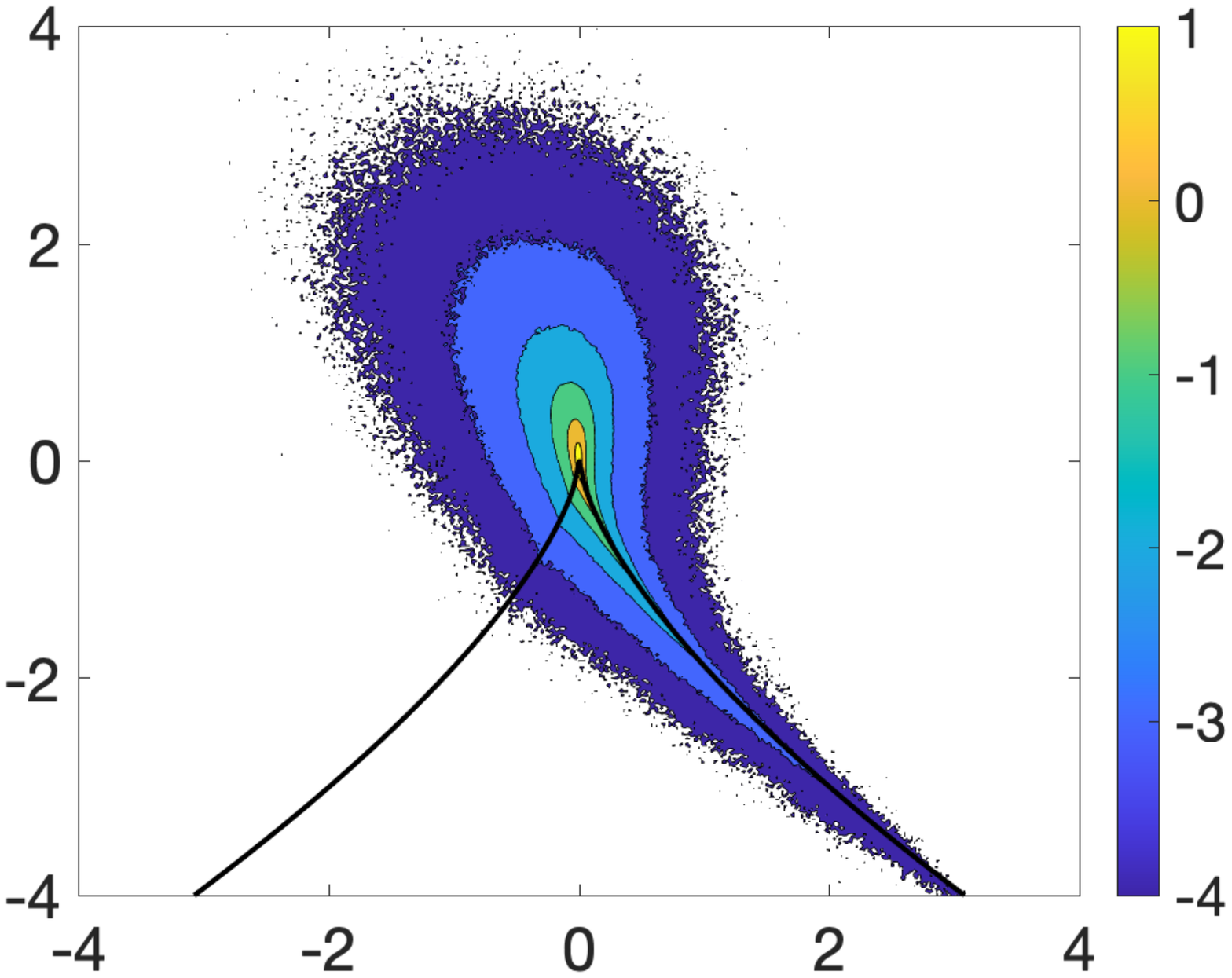}
				\put(75,-1){$R/\tilde{Q}_{av}^{3/2}$}
				\put(8,55){\rotatebox{90}{$Q/\tilde{Q}_{av}$}}								
		\end{overpic}}	
				\centering
		\subfloat[]
		{\begin{overpic}
				[trim = 0mm 60mm 10mm 60mm,scale=0.3,clip,tics=20]{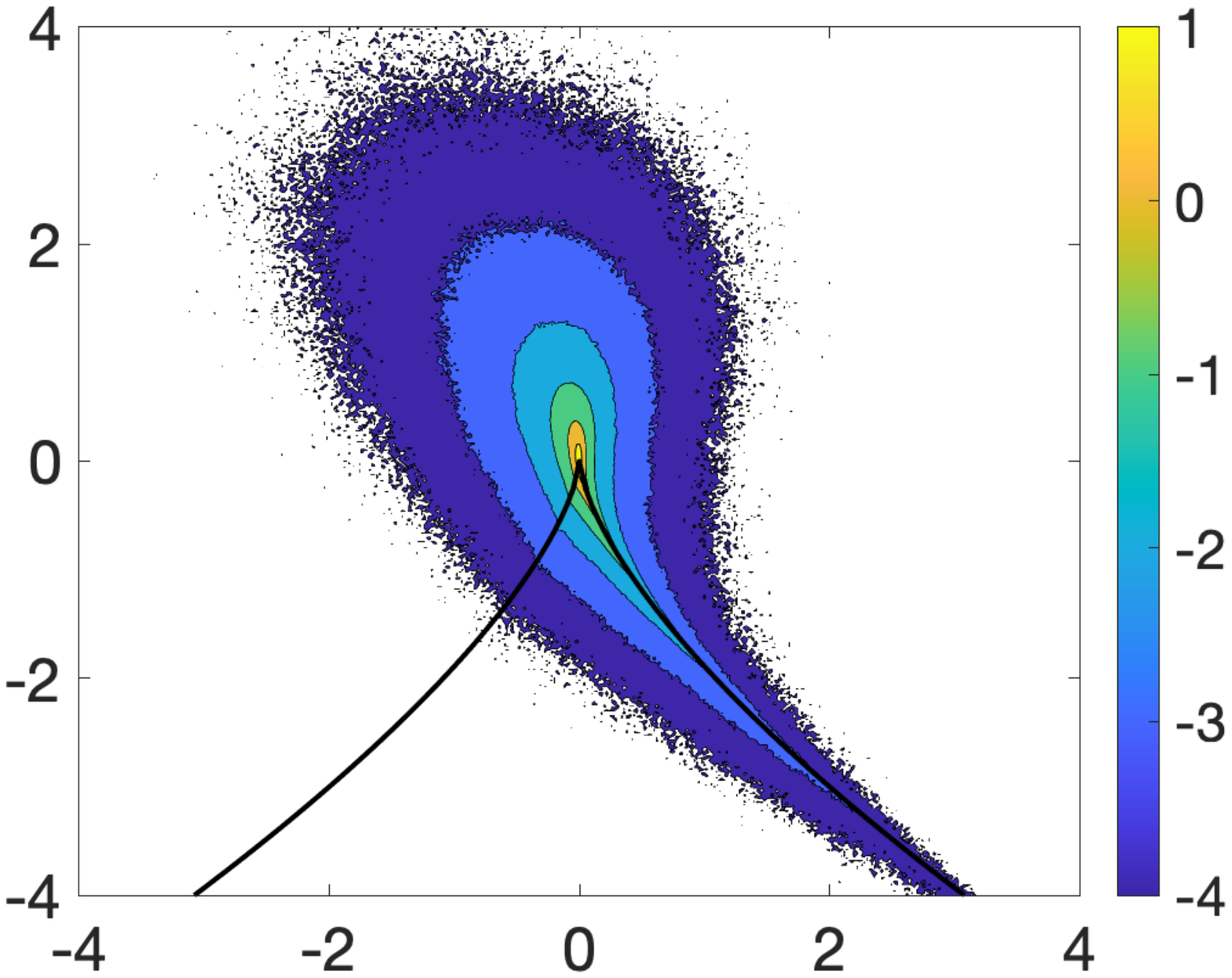}
				\put(75,-1){$R/\tilde{Q}_{av}^{3/2}$}
				\put(8,55){\rotatebox{90}{$Q/\tilde{Q}_{av}$}}							
		\end{overpic}}	
		\caption{Logarithm of joint-PDFs of $Q/\tilde{Q}_{av}$ and $R/\tilde{Q}_{av}^{3/2}$ (where $\tilde{Q}_{av}\equiv\langle\|\mathcal{A}\|^2\rangle$) from (a) DNS, (b) M1, (c) M2, (d) M3. Colours indicate the values of the logarithm of the PDF.} 
		\label{PDF_QR_A}
\end{figure}}

A more careful and quantitative comparison between the model predictions and DNS data for $Q,R$ can be made by comparing the PDFs of $Q$ and $R$ separately. In Figure \ref{PDF_QR}(a), one can see that in the DNS, the PDF of $Q$ is strongly positively skewed, which is associated with the fact that the vorticity is more intermittent than the strain-rate in turbulent flows \citep{yeung18}. By contrast, the ML-RDGF model predicts PDFs for $Q$ that are much more symmetric, with the model significantly underpredicting regions of intense vorticity, and slightly underpredicting regions of intense straining. These discrepancies are, however, mainly in the tails of the PDF, with the model predictions in good agreement with the DNS for values of the PDF that are $\geq O(10^{-2})$. 

In Figure \ref{PDF_QR}(b) it is seen that the model also predicts PDFs of $R$ that are relatively symmetric compared with the DNS data for which the PDF is negatively skewed. The model prediction for M2, which corresponds to $\Rey_\lambda=300$, is much closer to the DNS data than those of M1 which has the same $\Rey_\lambda=100$ as the DNS.

In view of these results, it is seen that while the ML-RDGF model predicts the components of the velocity gradients very accurately (as well as the alignments between the vorticity and strain-rate eigenvectors\citep{johnson2017turbulence}), it does not predict the invariants of the  velocity gradients accurately when compared with the DNS, at least not for the $\Rey_\lambda$ considered here. This could in turn lead to inaccuracies in the scalar gradient model, above and beyond any arising from the closure approximations for the scalar gradient diffusion term. These results point to the need for further refinements in the ML-RDGF model.

{\vspace{0mm}\begin{figure}
		\centering
		\subfloat[]
		{\begin{overpic}
				[trim = 5mm 60mm 0mm 60mm,scale=0.3,clip,tics=20]{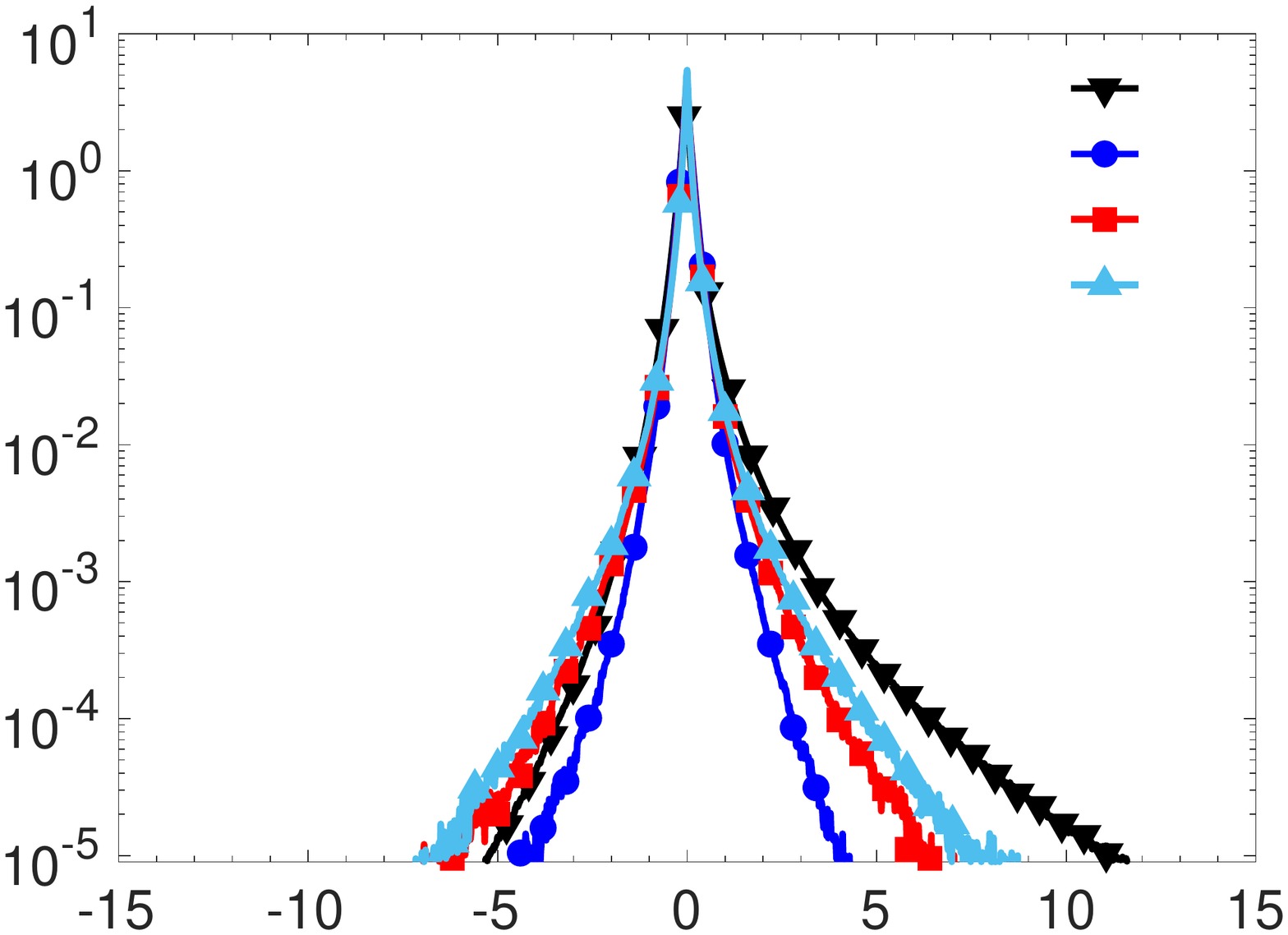}
				\put(80,-2){$Q/\tilde{Q}_{av}$}
				\put(-13,60){\rotatebox{90}{PDF}}
				\put(122,115){DNS}
				\put(122,107){M1}
				\put(122,98){M2}
			    \put(122,89){M3}				
		\end{overpic}}
		\centering
		\subfloat[]
		{\hspace{2mm}\begin{overpic}
				[trim = 5mm 60mm 0mm 60mm,scale=0.3,clip,tics=20]{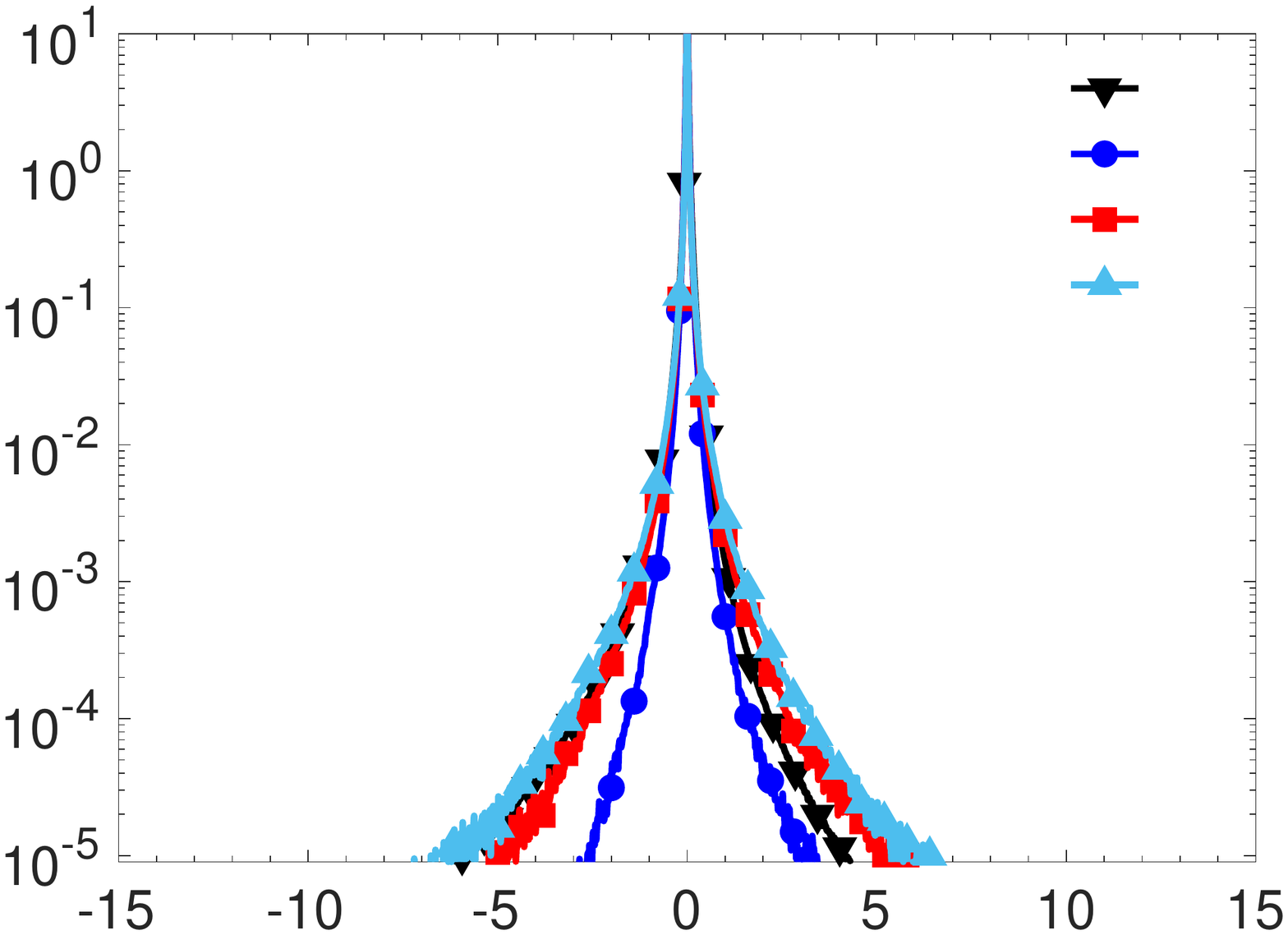}
				\put(80,-2){$R/\tilde{Q}_{av}^{3/2}$}
				\put(-13,60){\rotatebox{90}{PDF}}
				\put(122,115){DNS}
				\put(122,107){M1}
				\put(122,98){M2}
			    \put(122,89){M3}				
		\end{overpic}}		
		\caption{PDFs of (a) $Q/\tilde{Q}_{av}$ and (b) $R/\tilde{Q}_{av}^{3/2}$, where $\tilde{Q}_{av}\equiv\langle\|\boldsymbol{\mathcal{A}}\|^2\rangle$.} 
		\label{PDF_QR}
\end{figure}}

\subsection{\label{sec:level2_SG} Scalar Gradients}

We now turn to consider the predictions from our new model for the scalar gradients. In Figure \ref{PDF_B2}, we plot the PDFs of $Q_b/Q_{b,av}$ (where $Q_b\equiv\|\boldsymbol{b}\|^2$) which is proportional to the scalar dissipation rate $\epsilon_\phi\equiv \kappa\|\boldsymbol{b}\|^2$, as well as the PDF of $b_1$, the scalar gradient component in one of the (arbitrary, due to isotropy) directions. Concerning the PDF of $Q_b/Q_{b,av}$, the results show that while the model is in good qualitative agreement with the DNS data, capturing the slowly decaying tail of the PDF, it significantly underpredicts the values of the PDF. The model predictions for the PDF of $b_1$ are also in significant error, underpredicting small to intermediate values of $b_1/b_{1,rms}$, and significantly overpredicting large values of $b_1/b_{1,rms}$, such that the overall shape of the PDF is not well captured by the model. 

An investigation into the cause of these significant underpredictions revealed that the problem is due to the model generating extremely large values of $\|\boldsymbol{\mathcal{B}}(t)\|^2$. An example of the time series of $\|\boldsymbol{\mathcal{B}}(t)\|^2/\langle \|\boldsymbol{\mathcal{B}}(t)\|^2\rangle$ generated by the model at $\Rey_\lambda=100$ is shown in Figure \ref{time_series}, together with the time series of $\|\boldsymbol{\mathcal{S}}(t)\|^2/\langle \|\boldsymbol{\mathcal{S}}(t)\|^2\rangle$ for comparison. Although the signal $\|\boldsymbol{\mathcal{S}}(t)\|^2$ exhibits significant fluctuations about the mean, $\|\boldsymbol{\mathcal{B}}(t)\|^2$ exhibits infrequent but enormous fluctuations about the mean, which only get stronger as $\Rey_\lambda$ is increased. Although $\|\boldsymbol{\mathcal{B}}(t)\|^2$ would be expected to be more intermittent than $\|\boldsymbol{\mathcal{S}}(t)\|^2$, one would not anticipate intermittent fluctuations in $\|\boldsymbol{\mathcal{B}}(t)\|^2$ as large as these, nor are they manifested in the DNS data, and therefore they seem to indicate an issue with the model. Since the integral of the PDF of $Q_b$ over its sample-space is one, then because the model vastly overpredicts the probability of extremely large values of $\|\boldsymbol{\mathcal{B}}(t)\|^2$, it underpredicts the probability of values in the sample-space range shown in figure \ref{PDF_B2}.

{\vspace{0mm}\begin{figure}
		\centering
				\subfloat[]
		{\begin{overpic}
				[trim =5mm 60mm 0mm 60mm,scale=0.3,clip,tics=20]{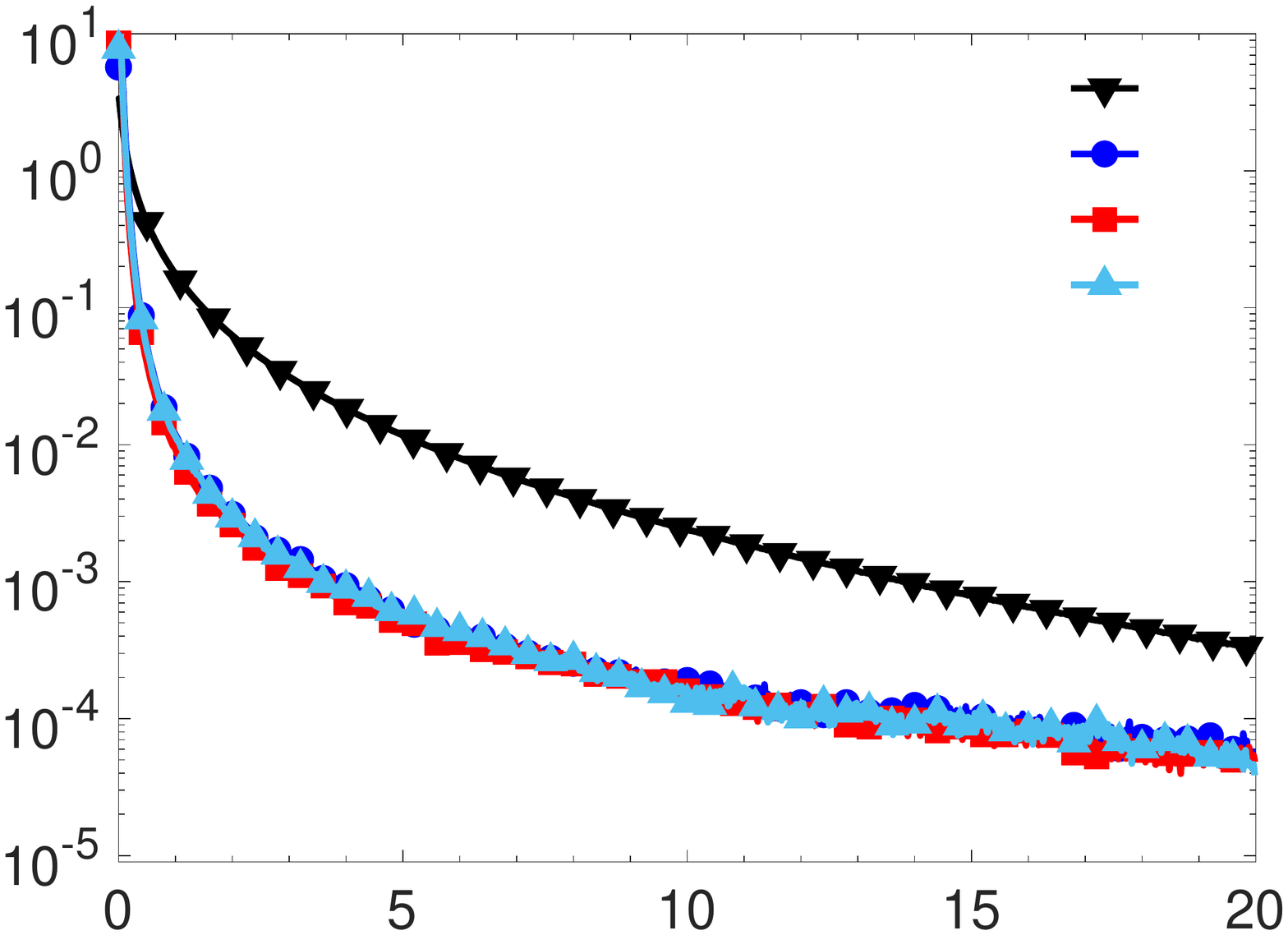}
				\put(80,-2){$Q_b/Q_{b,av}$}
				\put(-13,60){\rotatebox{90}{PDF}}
				\put(122,115){DNS}
				\put(122,107){M1}
				\put(122,98){M2}
			    \put(122,89){M3}			
		\end{overpic}}
				\subfloat[]
				{\hspace{2mm}\begin{overpic}
				[trim =5mm 60mm 0mm 60mm,scale=0.3,clip,tics=20]{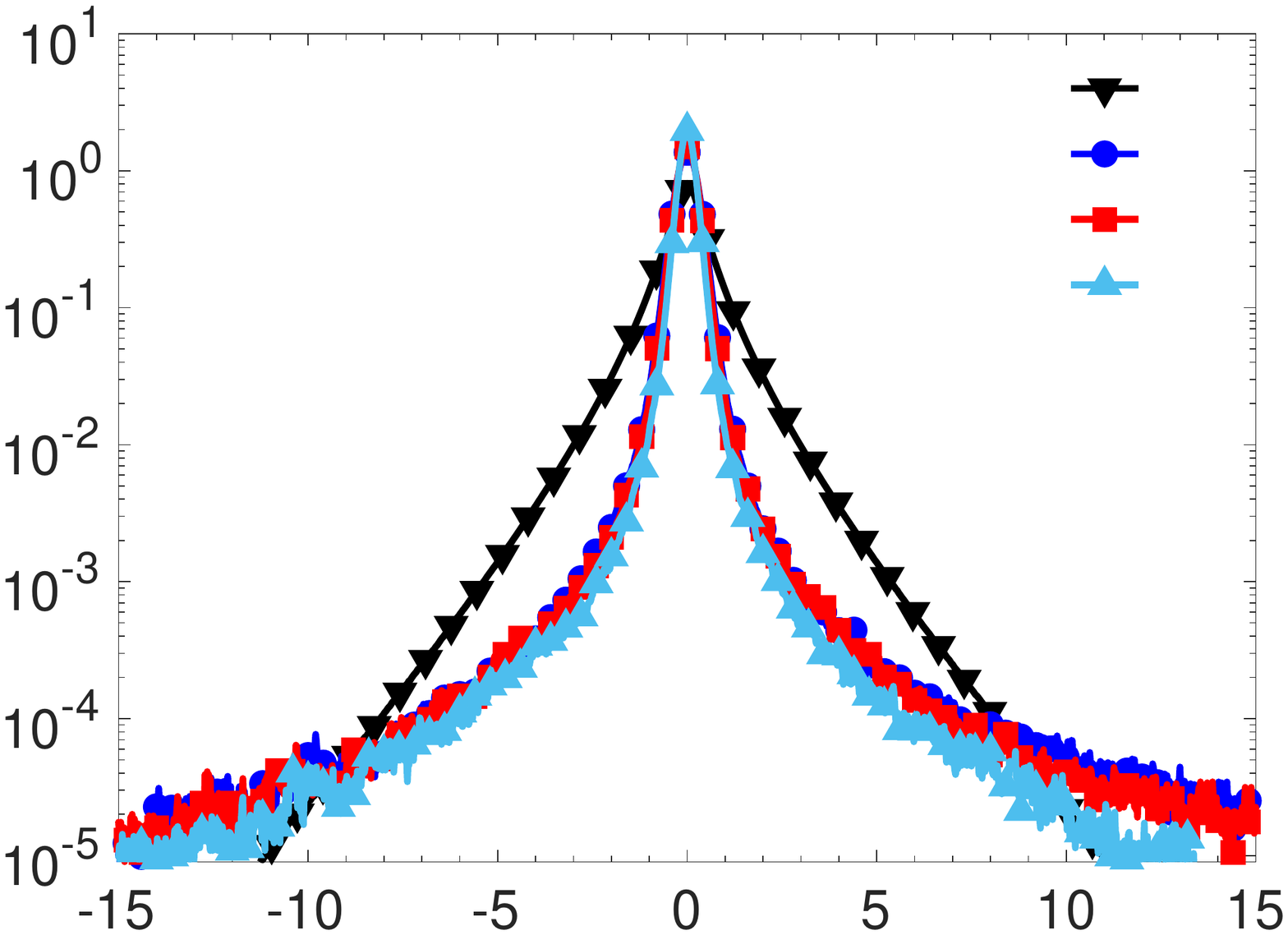}
				\put(80,-2){$b_1/b_{1,rms}$}
				\put(-13,60){\rotatebox{90}{PDF}}
				\put(122,115){DNS}
				\put(122,107){M1}
				\put(122,98){M2}
			    \put(122,89){M3}				
		\end{overpic}}
		\caption{PDFs of (a) $Q_b/Q_{b,av}$, (b) $b_1/b_{1,rms}$. The results from the model are obtained with the (uncorrected) model coefficient $\alpha_\mathcal{B}$ specified by equation \eqref{alphaB}.} 
		\label{PDF_B2}
\end{figure}}

{\vspace{0mm}\begin{figure}
		\centering
				\subfloat[]
		{\begin{overpic}
				[trim =5mm 60mm 0mm 60mm,scale=0.3,clip,tics=20]{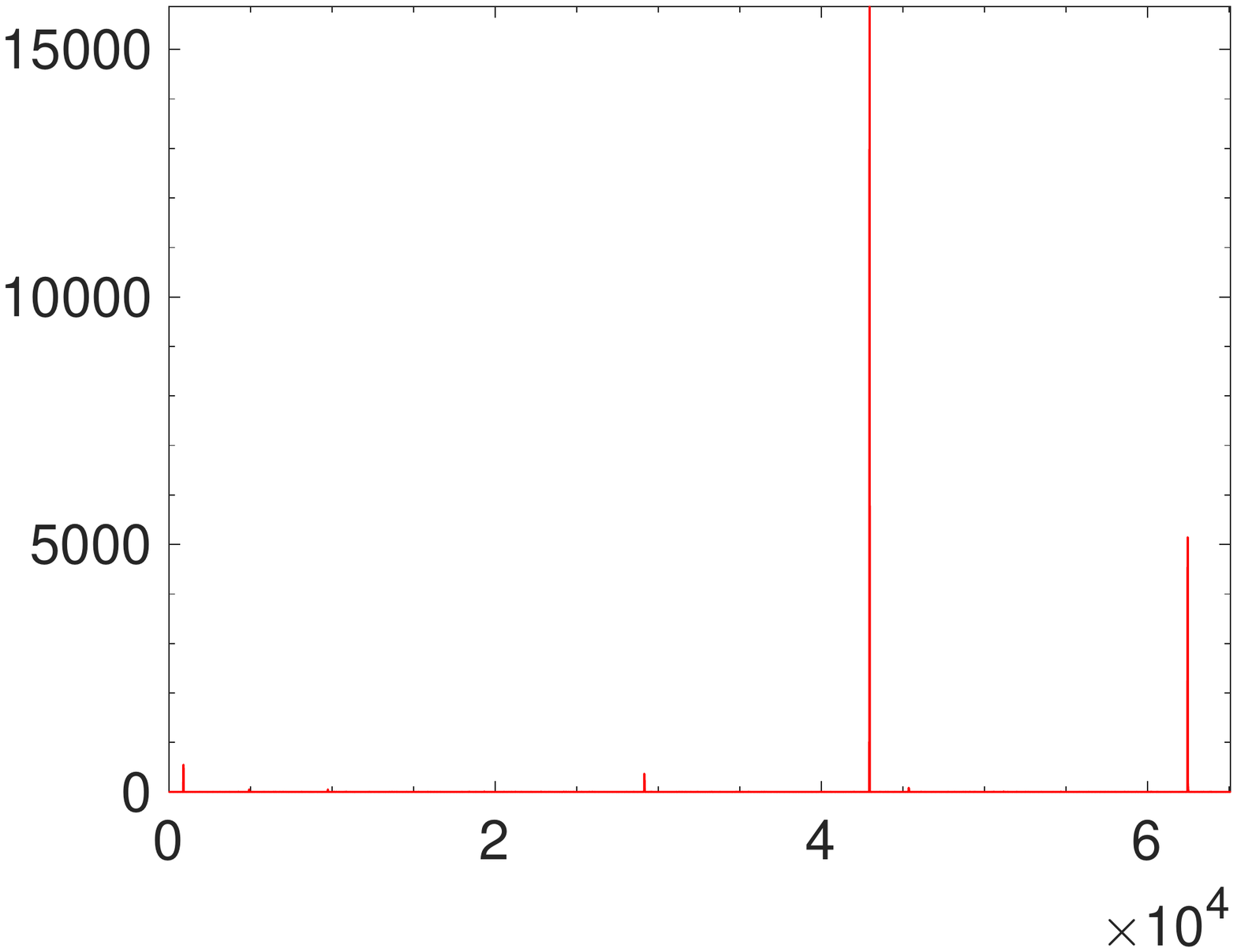}
				\put(80,2){$t/\tau_\eta$}
				\put(-15,40){\rotatebox{90}{$\|\boldsymbol{\mathcal{B}}(t)\|^2/\langle \|\boldsymbol{\mathcal{B}}(t)\|^2\rangle$}}
		\end{overpic}}
				\subfloat[]
		{\hspace{2mm}\begin{overpic}
				[trim =5mm 60mm 0mm 60mm,scale=0.3,clip,tics=20]{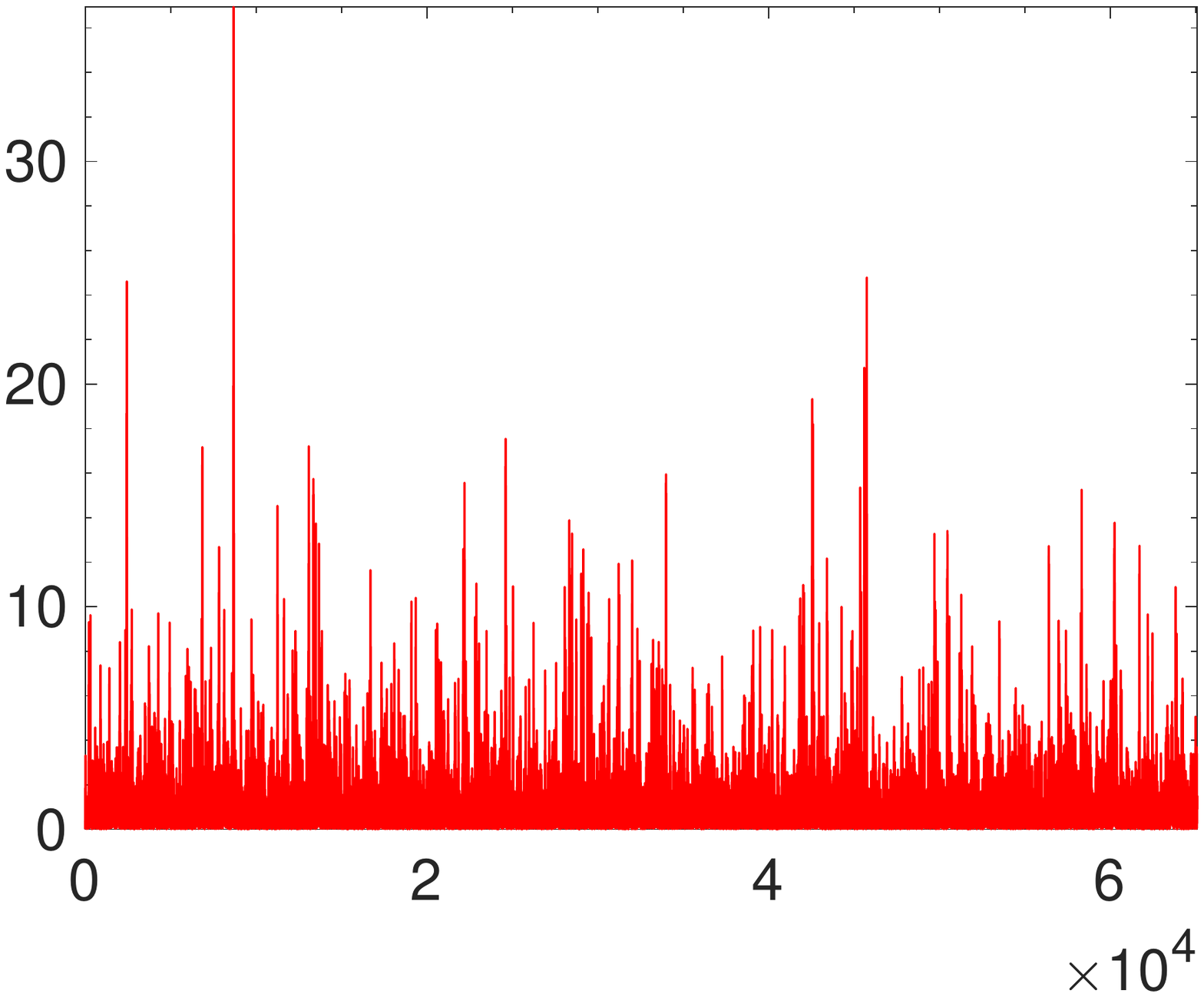}
				\put(80,2){$t/\tau_\eta$}
				\put(-5,40){\rotatebox{90}{$\|\boldsymbol{\mathcal{S}}(t)\|^2/\langle \|\boldsymbol{\mathcal{S}}(t)\|^2\rangle$}}
		\end{overpic}}		
		\caption{Time series of $\|\boldsymbol{\mathcal{B}}(t)\|^2$ and $\|\boldsymbol{\mathcal{S}}(t)\|^2$, normalized by their mean values, generated from the model with $\Rey_\lambda=100$ and the (uncorrected) coefficient $\alpha_\mathcal{B}$ specified by equation \eqref{alphaB}.} 		
		\label{time_series}
\end{figure}}

That the model vastly overpredicts the probability of extremely large values of $\|\boldsymbol{\mathcal{B}}(t)\|^2/\langle \|\boldsymbol{\mathcal{B}}(t)\|^2\rangle$ must be due to deficiencies in the closure for $\kappa\langle\nabla^2\boldsymbol{B}\rangle_{\boldsymbol{\mathcal{A}},\boldsymbol{\mathcal{B}}}$. In particular, the values of $\kappa\langle\nabla^2\boldsymbol{B}\rangle_{\boldsymbol{\mathcal{A}},\boldsymbol{\mathcal{B}}}$ predicted by the closure approximation when $\|\boldsymbol{\mathcal{B}}\|$ is large are too small to sufficiently counteract the scalar production term (although apparently, they are large enough to prevent singularities in the model since simulations of the model do not blow up). A relatively simple modification to the model to address this deficiency is to modify its specification of the coefficient $\alpha_\mathcal{B}$ in \eqref{alphaB} so that it includes information on the locally averaged scalar production, rather than simply the global mean value. This is achieved by replacing \eqref{alphaB} with
\begin{align}
\alpha_\mathcal{B}(t)&= \tau_\eta\int^t_{t-\tau_1} \boldsymbol{\mathcal{S}}(t')\boldsymbol{:}\boldsymbol{\mathcal{B}}(t')\boldsymbol{\mathcal{B}}(t')\,dt'\Bigg/\int^t_{t-\tau_1} \|\boldsymbol{\mathcal{B}}(t')\|^2\,dt',\label{alphaBnew}
\end{align}
which can be computed when solving the model since it only depends on $\boldsymbol{\mathcal{S}}$ and $\boldsymbol{\mathcal{B}}$ at previous times, which are known. With this, the global average involved in \eqref{alphaB} is replaced with a local time average over the trajectory history of the particle. The time integral is chosen to span $[t-\tau_1,t]$ in view of the fact that $\boldsymbol{\mathcal{S}}\boldsymbol{:}\boldsymbol{\mathcal{B}}\boldsymbol{\mathcal{B}}$ and $\|\boldsymbol{\mathcal{B}}\|^2$ have timescales on the order of the integral timescale, which in the ML-RDGF is specified by $\tau_1$. 

The advantage of using \eqref{alphaBnew} is that the coefficient $\delta_\mathcal{B}$ in \eqref{Beq_closed_forcing} will then depend upon the local scalar gradient dynamics, and in regions where the production of $\|\boldsymbol{\mathcal{B}}\|^2$ is large, $\delta_\mathcal{B}$ will also be large relative to its value in regions where the production of $\|\boldsymbol{\mathcal{B}}\|^2$ is small. In other words, using \eqref{alphaBnew} introduces nonlinearity into the closure for $\kappa\langle\nabla^2\boldsymbol{B}\rangle_{\boldsymbol{\mathcal{A}},\boldsymbol{\mathcal{B}}}$ with respect to its dependence on $\boldsymbol{\mathcal{B}}$, and this may help oppose the extremely large fluctuations predicted by the original form of the model. In practice, since \eqref{alphaBnew} requires time-history information, the model is solved for $t\leq \tau_1$ using \eqref{alphaB}, and then for $t>\tau_1$, \eqref{alphaBnew} is used.

{\vspace{0mm}\begin{figure}
		\centering
				\subfloat[]
		{\begin{overpic}
				[trim =5mm 60mm 0mm 60mm,scale=0.3,clip,tics=20]{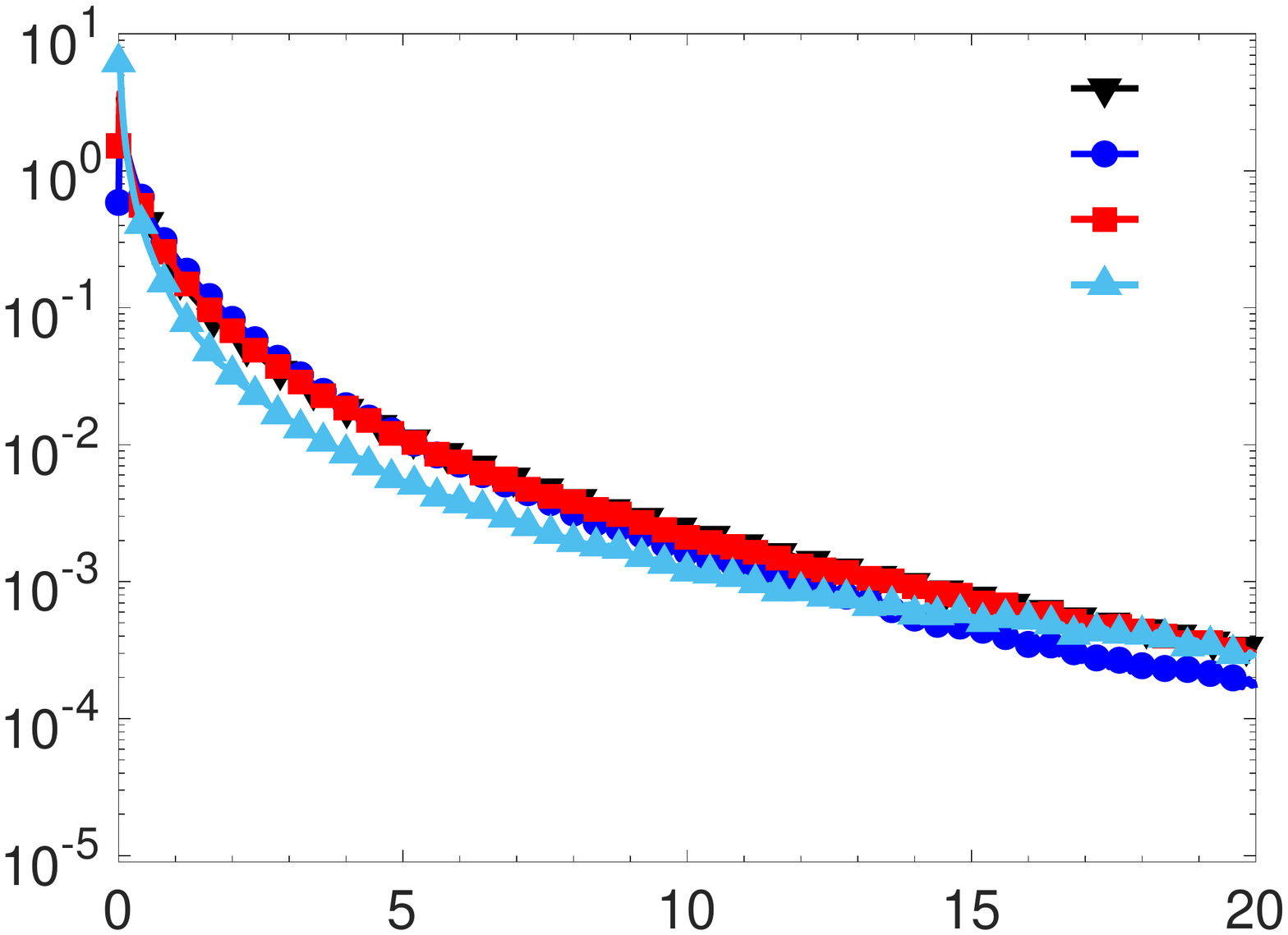}
				\put(80,-2){$Q_b/Q_{b,av}$}
				\put(-13,60){\rotatebox{90}{PDF}}
				\put(122,115){DNS}
				\put(122,107){M1}
				\put(122,98){M2}
			    \put(122,89){M3}				
		\end{overpic}}
				\subfloat[]
				{\hspace{2mm}\begin{overpic}
				[trim =5mm 60mm 0mm 60mm,scale=0.3,clip,tics=20]{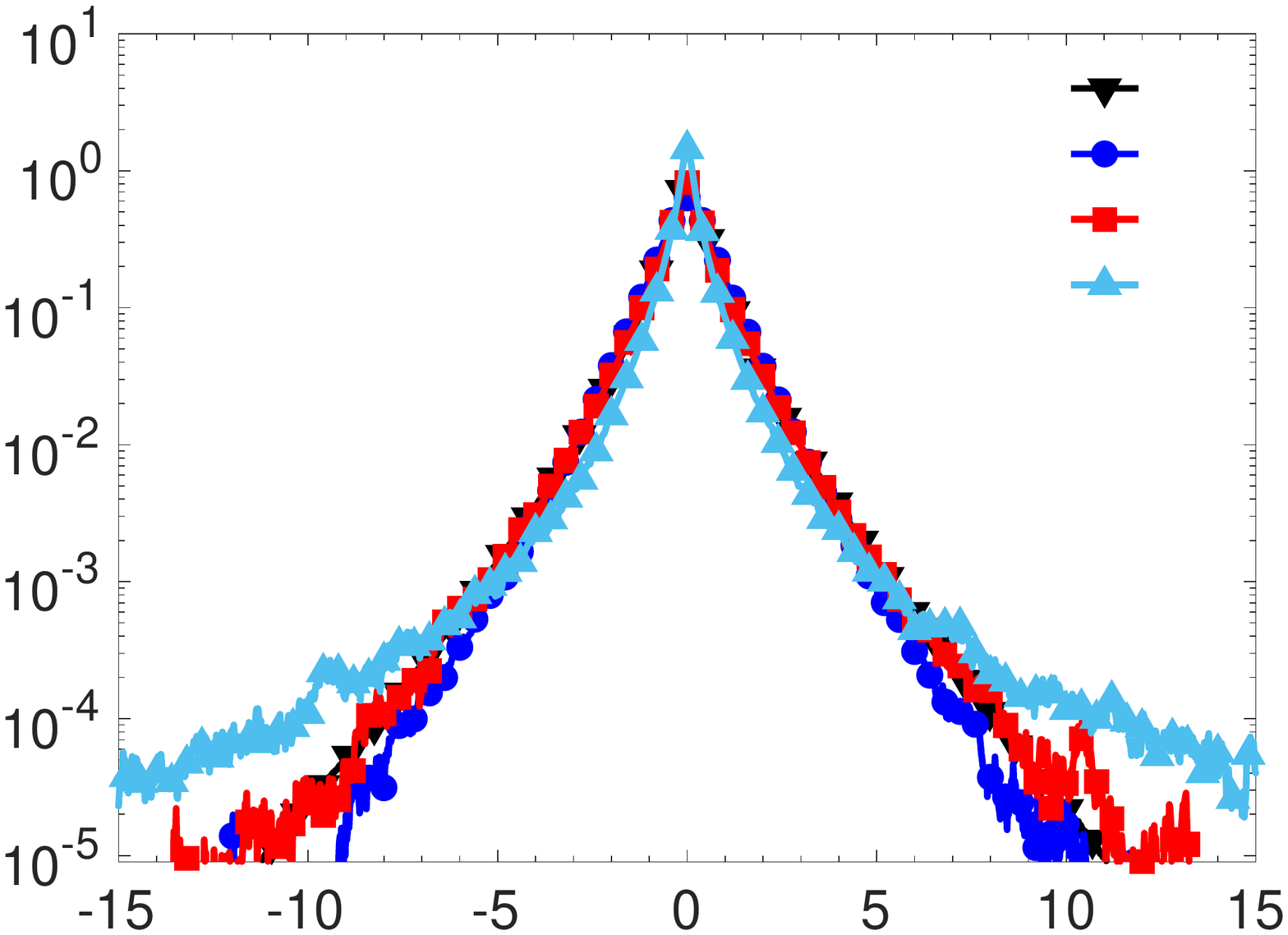}
				\put(80,-2){$b_1/b_{1,rms}$}
				\put(-13,60){\rotatebox{90}{PDF}}
				\put(122,115){DNS}
				\put(122,107){M1}
				\put(122,98){M2}
			    \put(122,89){M3}				
		\end{overpic}}
		\caption{PDFs of (a) $Q_b/Q_{b,av}$, (b) $b_1/b_{1,rms}$, based on using \eqref{alphaBnew} instead of \eqref{alphaB} to specify $\alpha_\mathcal{B}$ in the model.} 		
		\label{PDF_B2_new}
\end{figure}}

{\vspace{0mm}\begin{figure}
		\centering
				\subfloat[]
		{\begin{overpic}
				[trim =5mm 60mm 0mm 60mm,scale=0.3,clip,tics=20]{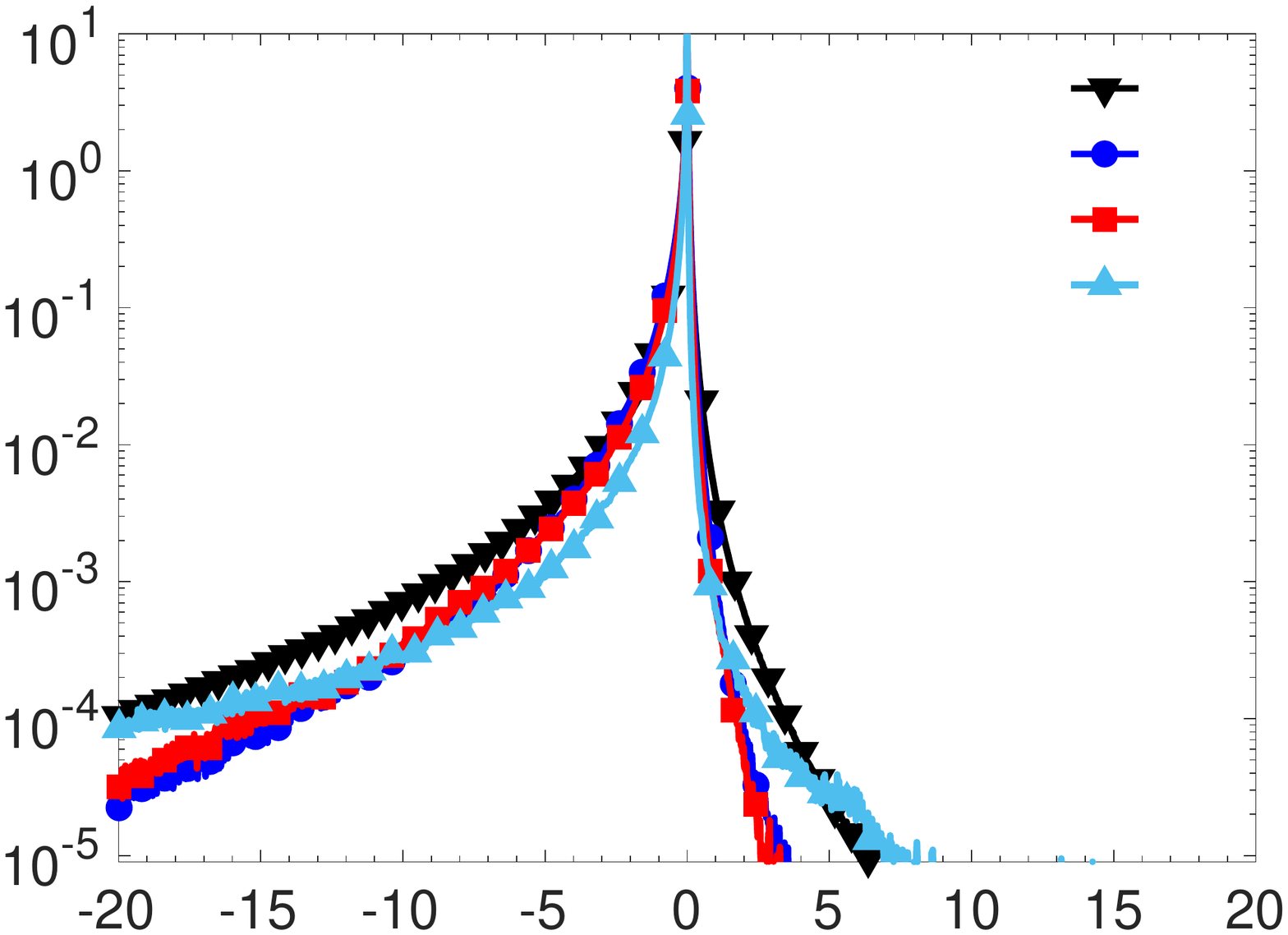}
				\put(73,-2){$R_b/\tilde{Q}_{av}^{1/2}Q_{b,av}$}
				\put(-10,60){\rotatebox{90}{PDF}}
				\put(122,115){DNS}
				\put(122,107){M1}
				\put(122,98){M2}
			    \put(122,89){M3}				
		\end{overpic}}
				\subfloat[]
				{\hspace{2mm}\begin{overpic}
				[trim =5mm 60mm 0mm 60mm,scale=0.3,clip,tics=20]{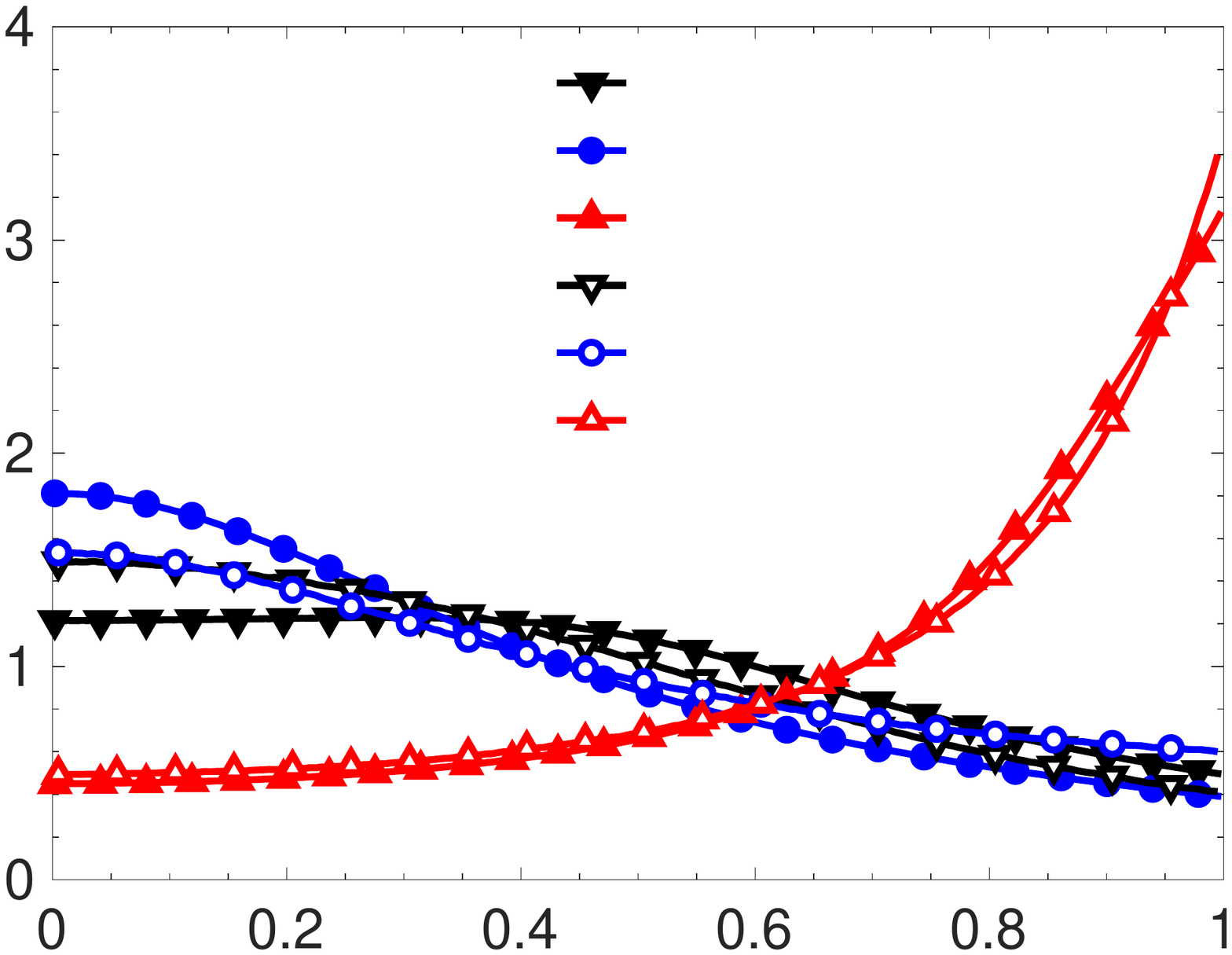}
				\put(80,-2){$\boldsymbol{e}_b\boldsymbol{\cdot}\boldsymbol{e}_i$}
				\put(-10,60){\rotatebox{90}{PDF}}
				\put(40,116){DNS, $i=1$}
				\put(40,107){DNS, $i=2$}
				\put(40,98){DNS, $i=3$}				
				\put(40,90){M1, $i=1$}
				\put(40,81){M1, $i=2$}
				\put(40,72){M1, $i=3$}				
		\end{overpic}}
		\caption{PDFs of (a) $R_b/\tilde{Q}_{av}^{1/2}Q_{b,av}$, (b) the inner product between the unit vector $\boldsymbol{e}_b\equiv\boldsymbol{b}/\|\boldsymbol{b}\|$ and the eigenvectors $\boldsymbol{e}_i$ of $\boldsymbol{\mathcal{S}}$.} 		
		\label{PDF_x4_align}
\end{figure}}

Figure \ref{PDF_B2_new} once again shows the PDFs of $Q_b$ and $b_1$, but this time using \eqref{alphaBnew} instead of \eqref{alphaB} to specify $\alpha_\mathcal{B}$ in the model. Comparing the results to those in figure \ref{PDF_B2}, it can be seen that the new specification of $\alpha_\mathcal{B}$ dramatically improves the predictions from the model, being now in excellent agreement with the DNS data. The results also show the impact of $\Rey_\lambda$ on $Q_b$ as predicted by the model, with the probability of intermediate values of $Q_b/Q_{b,av}$ and $b_1/b_{1,rms}$ predicted to decrease as $\Rey_\lambda$ increases, while the probability of large $Q_b/Q_{b,av}$ and $b_1/b_{1,rms}$ increases as $\Rey_\lambda$ increases. However, further tests of the model revealed that when $\Rey_\lambda$ is increased much beyond $\Rey_\lambda=500$ the predictions of the model become unrealistic, with extremely large values of the scalar gradient occurring in the model, and the model can even blow up. Therefore, although the use of \eqref{alphaBnew} dramatically improves the performance of the model over the range of $\Rey_\lambda$ considered, it is not sufficient to guarantee that the model makes reasonable predictions for arbitrarily large $\Rey_\lambda$. An investigation into the causes of the failure of the model at high $\Rey_\lambda$ and possible remedies for this are left to future work.

{\vspace{0mm}\begin{figure}
		\centering
		\subfloat[]
		{\begin{overpic}
				[trim = 0mm 60mm 10mm 60mm,scale=0.3,clip,tics=20]{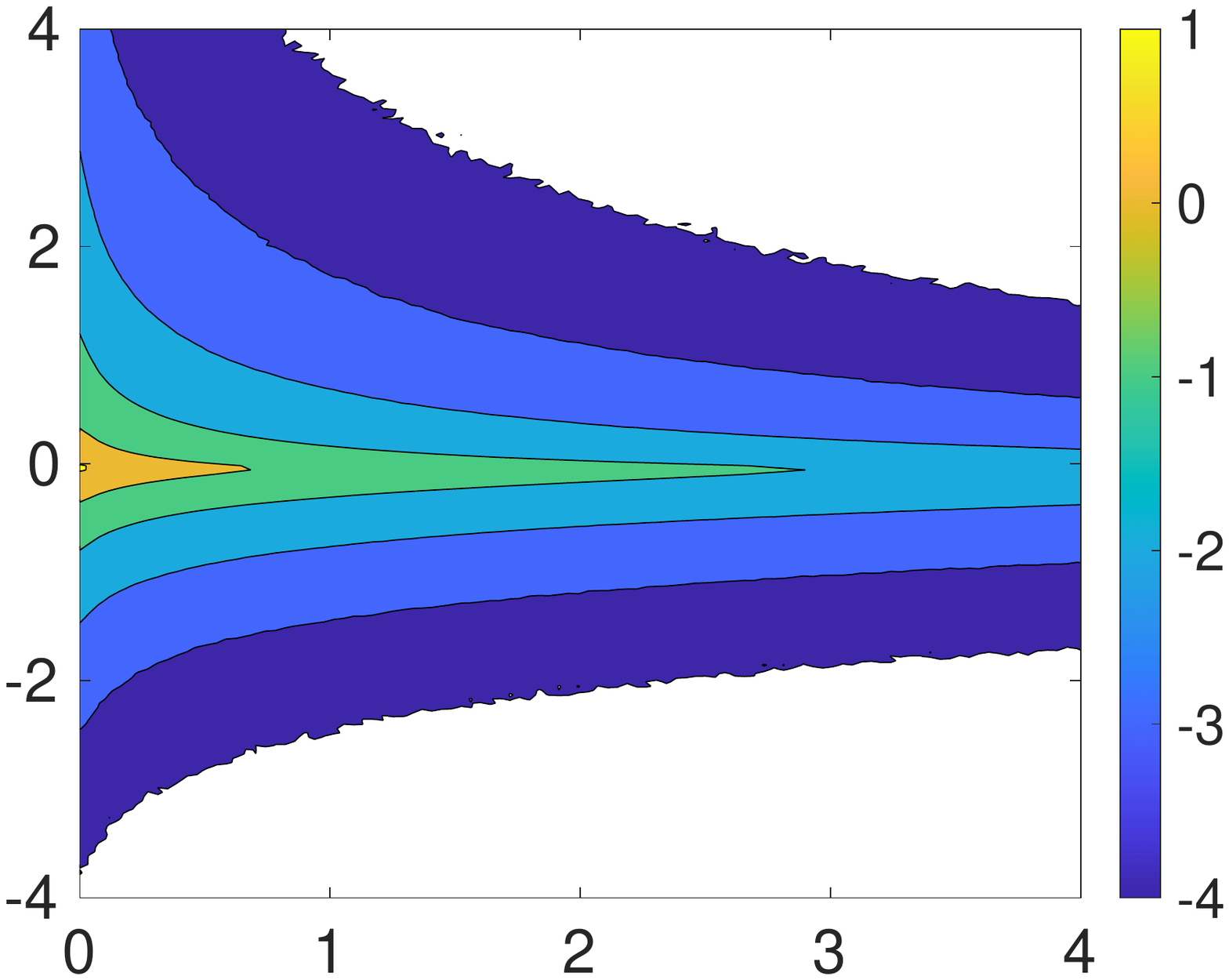}
				\put(75,-1){$Q_b/Q_{b,av}$}
				\put(8,55){\rotatebox{90}{$Q/\tilde{Q}_{av}$}}					
		\end{overpic}}
		\centering
		\subfloat[]
		{\begin{overpic}
				[trim = 0mm 60mm 10mm 60mm,scale=0.3,clip,tics=20]{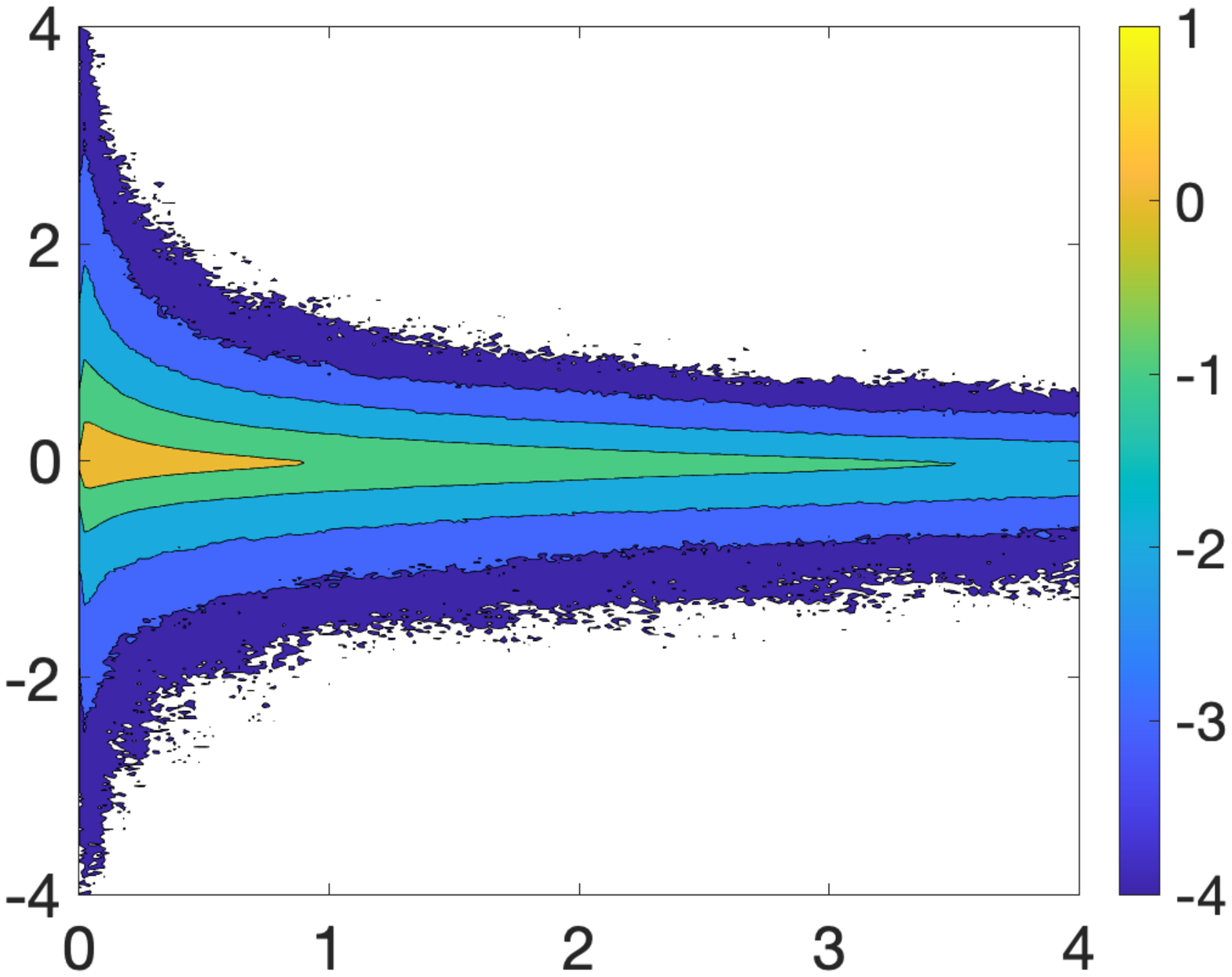}
				\put(75,-1){$Q_b/Q_{b,av}$}
				\put(8,55){\rotatebox{90}{$Q/\tilde{Q}_{av}$}}						
		\end{overpic}}\\
		\centering
		\subfloat[]
		{\begin{overpic}
				[trim = 0mm 60mm 10mm 60mm,scale=0.3,clip,tics=20]{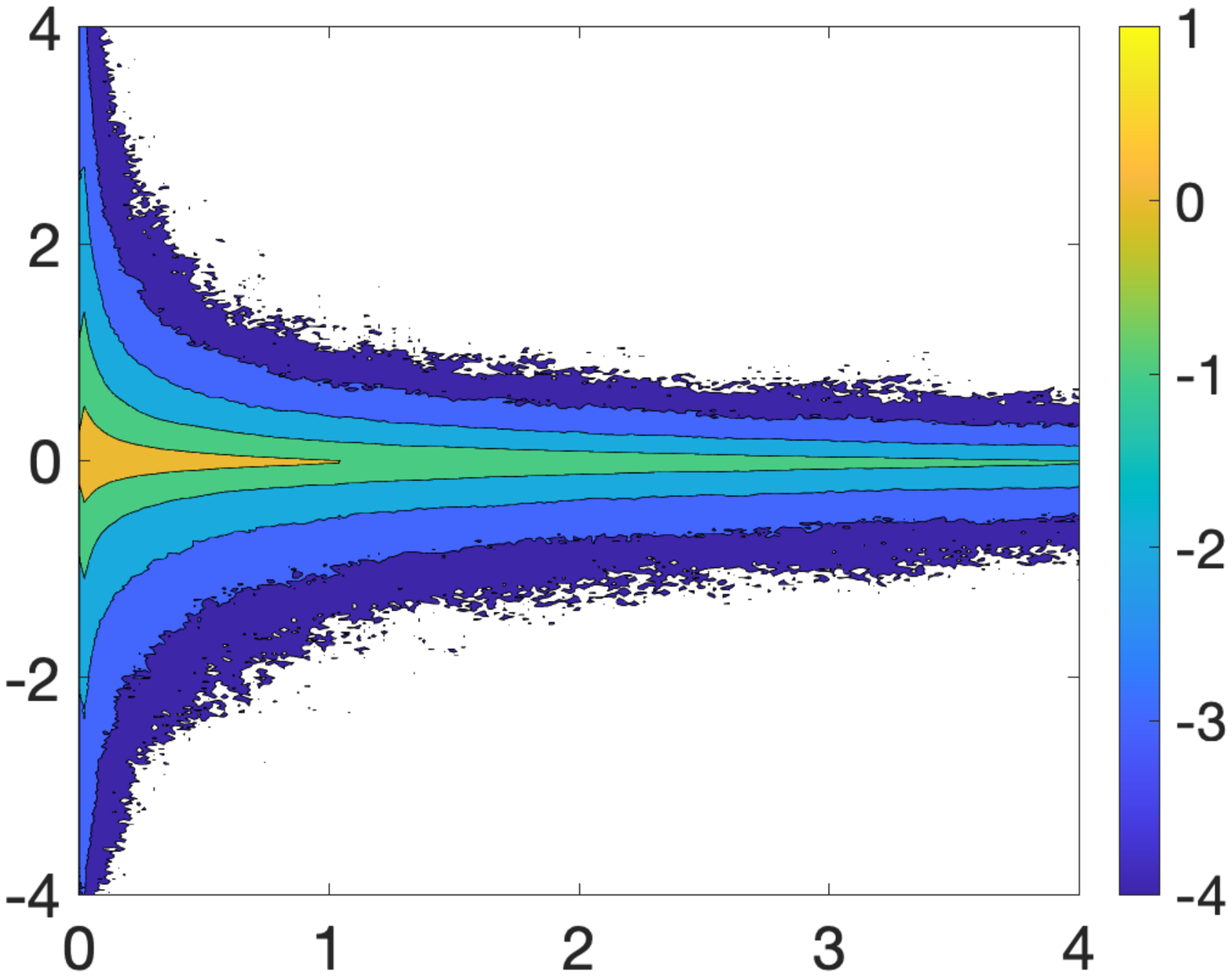}
				\put(75,-1){$Q_b/Q_{b,av}$}
				\put(8,55){\rotatebox{90}{$Q/\tilde{Q}_{av}$}}						
		\end{overpic}}	
				\centering
		\subfloat[]
		{\begin{overpic}
				[trim = 0mm 60mm 10mm 60mm,scale=0.3,clip,tics=20]{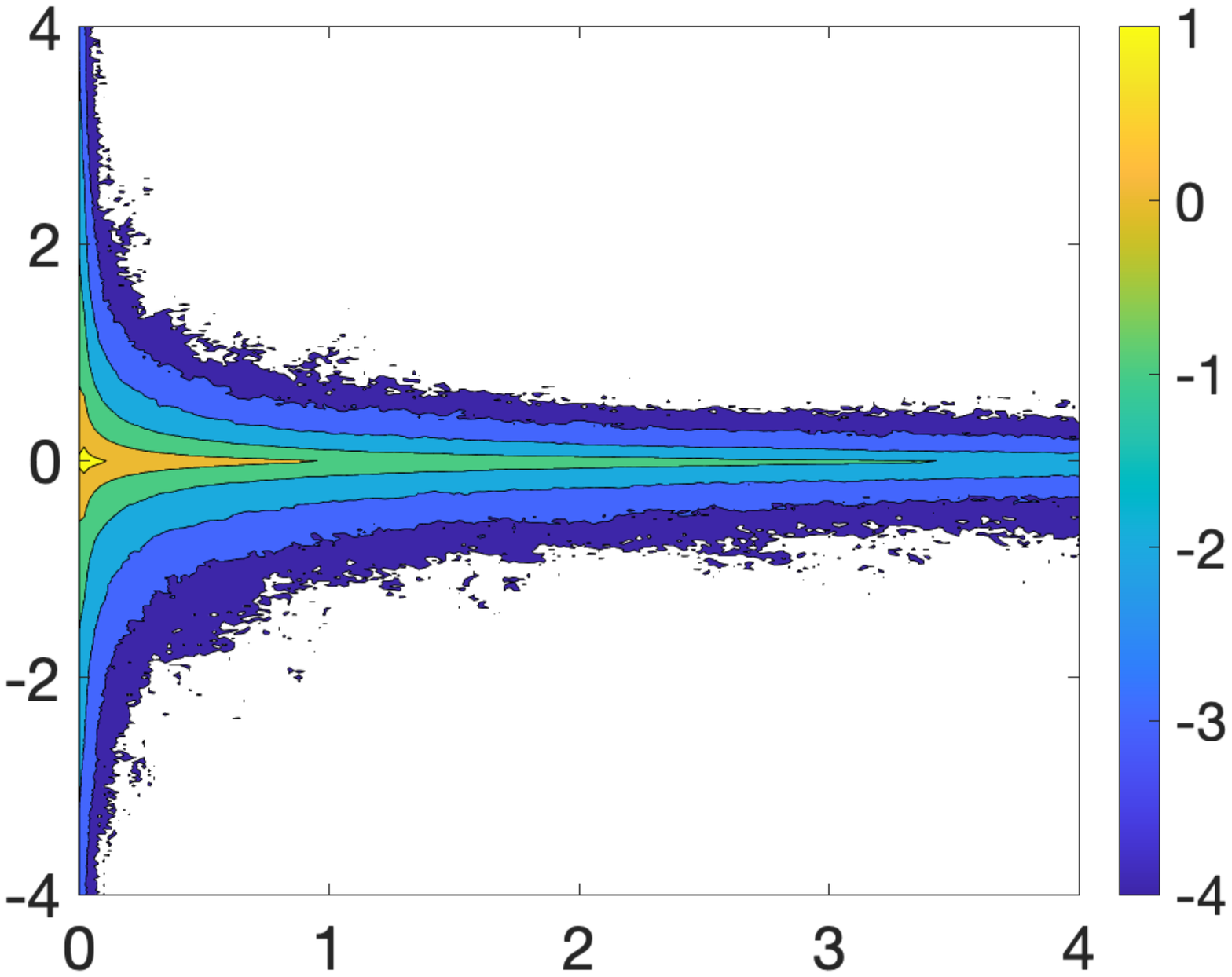}
				\put(75,-1){$Q_b/Q_{b,av}$}
				\put(8,55){\rotatebox{90}{$Q/\tilde{Q}_{av}$}}						
		\end{overpic}}	
		\caption{Logarithm of joint-PDFs of $Q/\tilde{Q}_{av}$ and $Q_b/Q_{b,av}$ from (a) DNS, (b) M1, (c) M2, (d) M3. Colors indicate the values of the logarithm of the PDF.} 
		\label{PDF_Q_Qb}
\end{figure}}

In figure \ref{PDF_x4_align}(a) we show the PDF of the scalar gradient production $R_b\equiv \boldsymbol{a:bb}$. The results show that the model predictions are in good agreement with the DNS data, with some underpredictions for the largest fluctuations. The model captures the strong negative skewness of the PDF that is associated with the predominance of scalar gradient production over destruction. The model also predicts the largest fluctuations in $R_b$ become more probable as $\Rey_\lambda$ is increased due to intermittency in the flow. In figure \ref{PDF_x4_align}(b) we show the PDF of the inner product between the unit vector $\boldsymbol{e}_b\equiv\boldsymbol{b}/\|\boldsymbol{b}\|$ and the eigenvectors $\boldsymbol{e}_i$ (corresponding to the ordered eigenvalues) of $\boldsymbol{\mathcal{S}}$. The model predicts these non-trivial alignments very well, capturing the strong preferential alignment with the compressional eigendirection $\boldsymbol{e}_3$, and misalignment with the intermediate eigendirection $\boldsymbol{e}_2$ and extensional eigendirection $\boldsymbol{e}_1$. However, the model predicts a misalignment with $\boldsymbol{e}_1$ that is a little too strong, and a misalignment with $\boldsymbol{e}_2$ that is a little too weak. Only the results for M1 are shown, as the results from M2 and M3 are almost identical. The current model predictions for these alignment PDFs are in much better agreement with the DNS data than those of the model of \cite{martin2005joint} which uses a much more simplistic closure approximation for the scalar diffusion term.

\subsection{\label{sec:level2_JPDF}Joint-PDFs of Velocity and Scalar Gradients}

{\vspace{0mm}\begin{figure}
		\centering
		\subfloat[]
		{\begin{overpic}
				[trim = 0mm 60mm 10mm 60mm,scale=0.3,clip,tics=20]{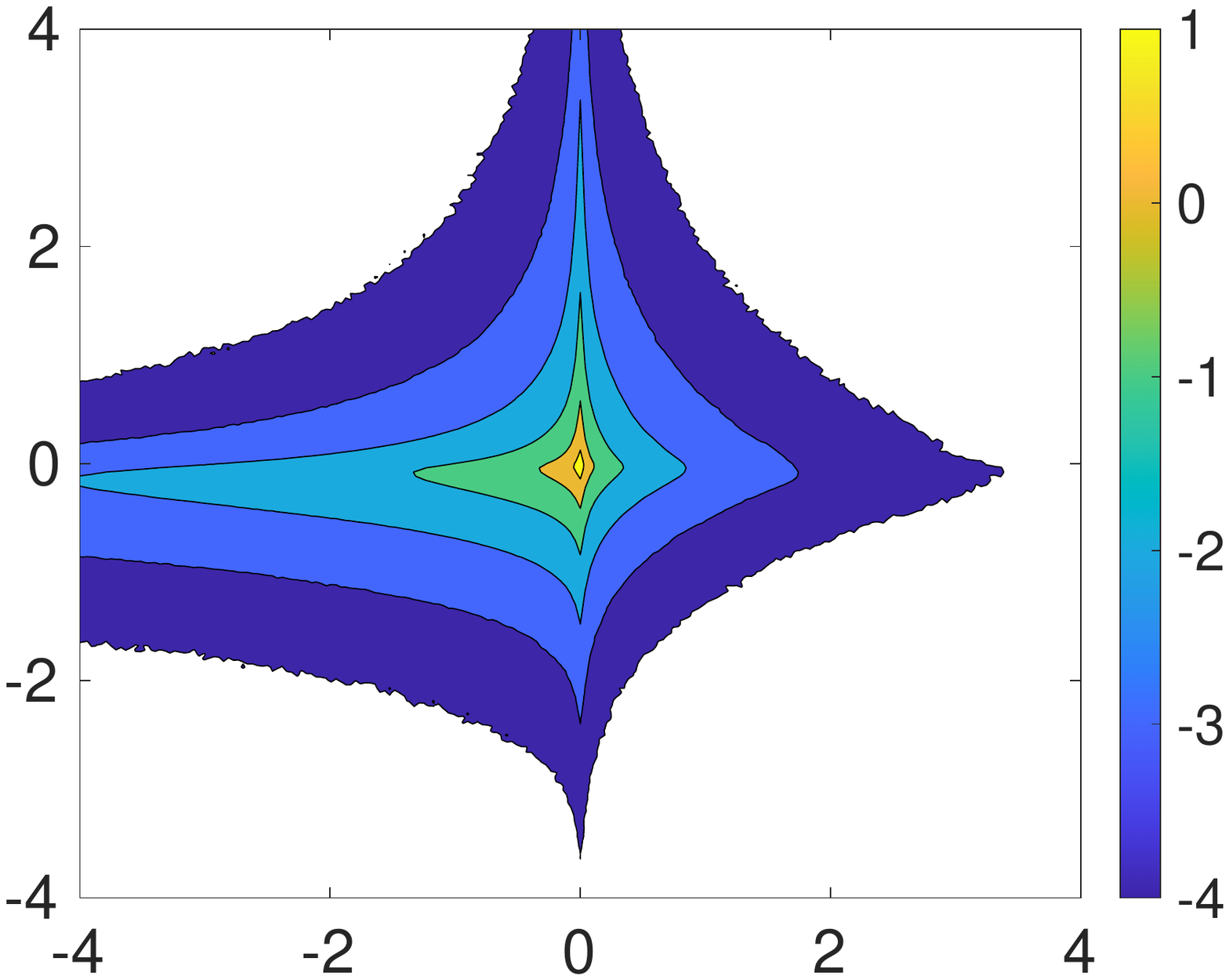}
				\put(70,-1){$R_b/\tilde{Q}_{av}^{1/2}Q_{b,av}$}
				\put(8,55){\rotatebox{90}{$Q/\tilde{Q}_{av}$}}					
		\end{overpic}}
		\centering
		\subfloat[]
		{\begin{overpic}
				[trim = 0mm 60mm 10mm 60mm,scale=0.3,clip,tics=20]{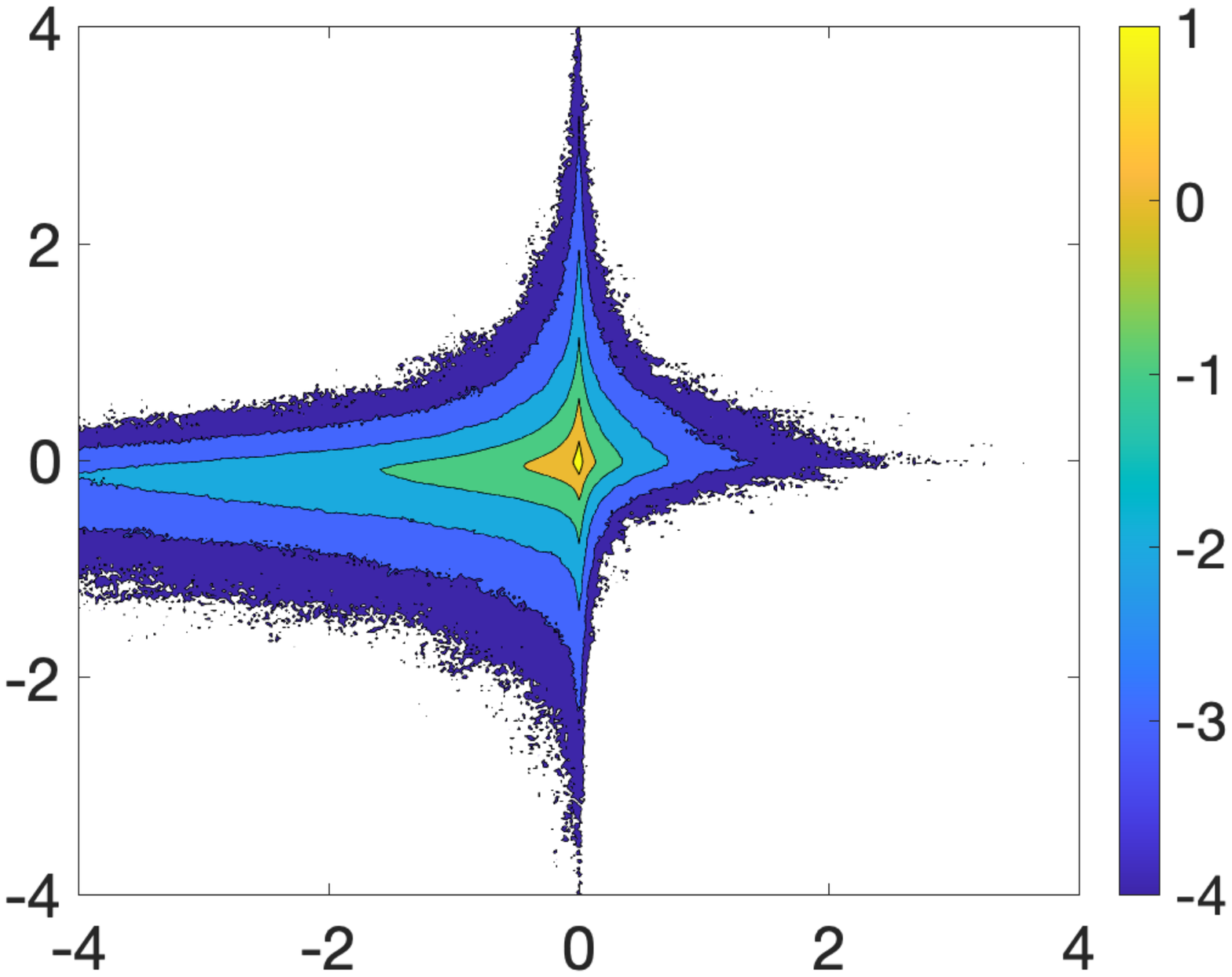}
				\put(70,-1){$R_b/\tilde{Q}_{av}^{1/2}Q_{b,av}$}
				\put(8,55){\rotatebox{90}{$Q/\tilde{Q}_{av}$}}			
		\end{overpic}}\\
		\centering
		\subfloat[]
		{\begin{overpic}
				[trim = 0mm 60mm 10mm 60mm,scale=0.3,clip,tics=20]{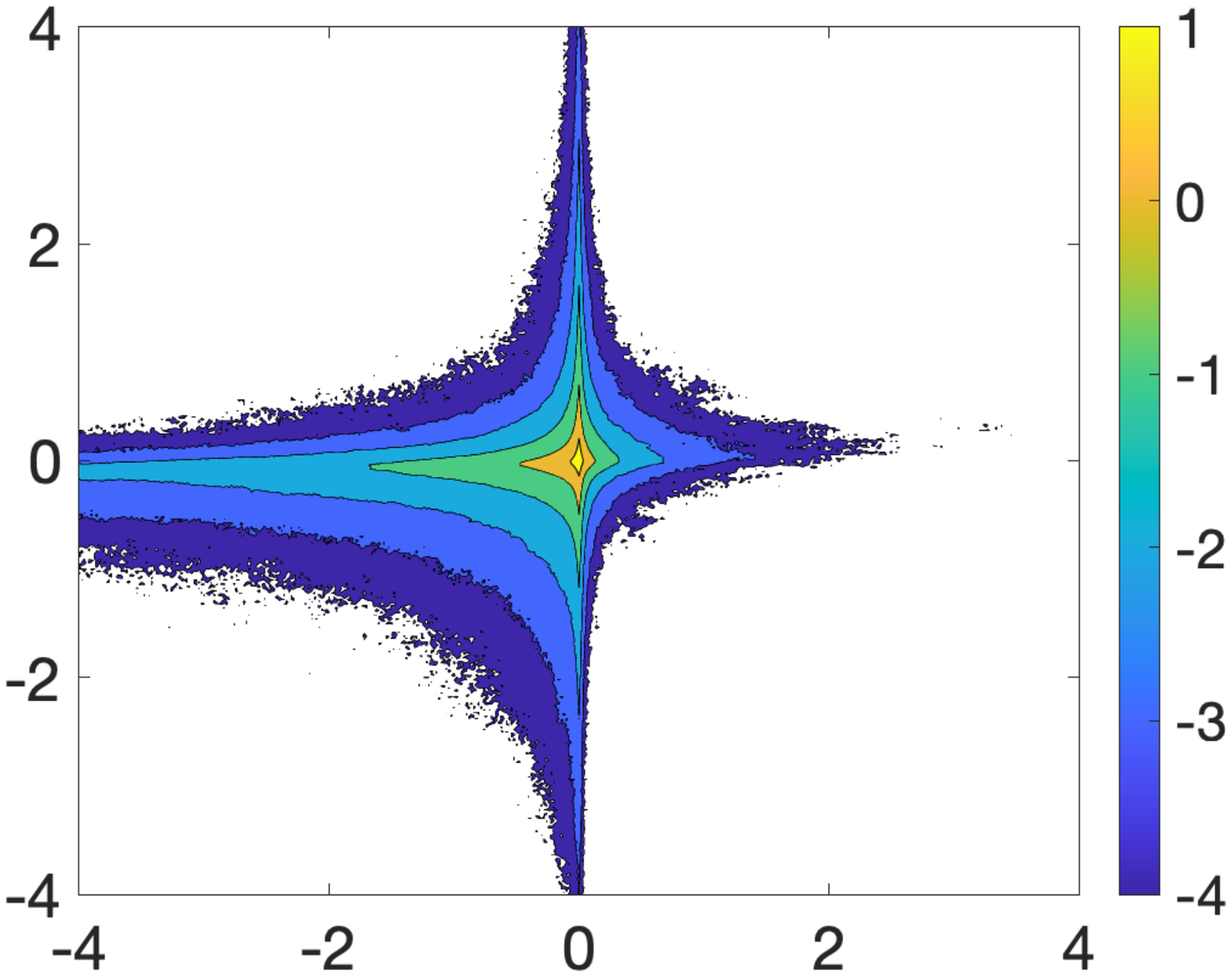}
				\put(70,-1){$R_b/\tilde{Q}_{av}^{1/2}Q_{b,av}$}
				\put(8,55){\rotatebox{90}{$Q/\tilde{Q}_{av}$}}			
		\end{overpic}}	
				\centering
		\subfloat[]
		{\begin{overpic}
				[trim = 0mm 60mm 10mm 60mm,scale=0.3,clip,tics=20]{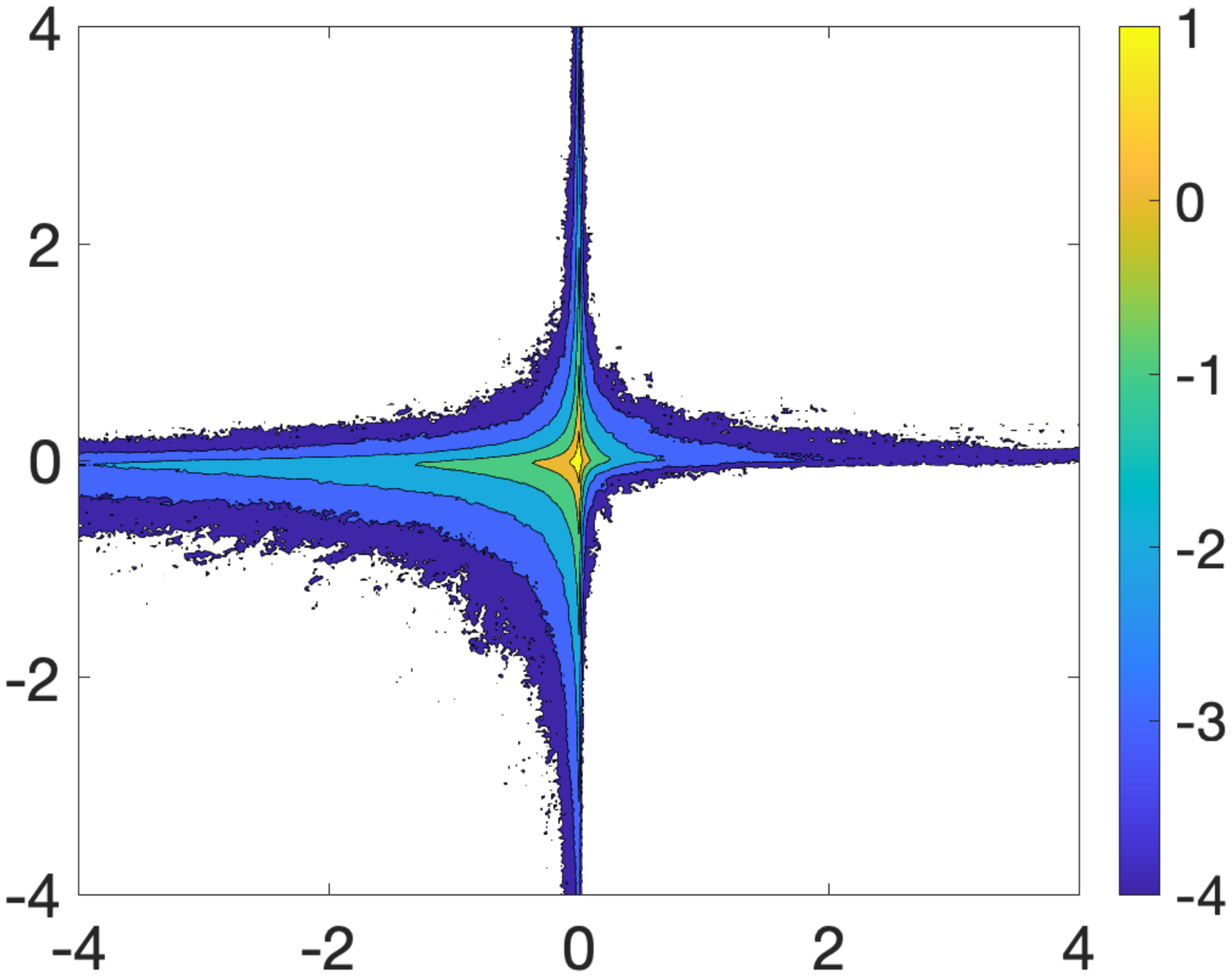}
				\put(70,-1){$R_b/\tilde{Q}_{av}^{1/2}Q_{b,av}$}
				\put(8,55){\rotatebox{90}{$Q/\tilde{Q}_{av}$}}				
		\end{overpic}}	
		\caption{Logarithm of joint-PDFs of $Q/\tilde{Q}_{av}$ and $R_b/\tilde{Q}_{av}^{1/2}Q_{b,av}$ from (a) DNS, (b) M1, (c) M2, (d) M3. Colors indicate the values of the logarithm of the PDF.} 
		\label{PDF_Q_Rb}
\end{figure}}

We now turn to consider the relationship between the velocity and scalar gradients predicted by the model. In Figure \ref{PDF_Q_Qb} we show the joint-PDFs of the velocity gradient invariant $Q/\tilde{Q}_{av}$ and scalar invariant $Q_b/Q_{b,av}$ from the DNS and model. For M1, there is an excellent qualitative agreement with the DNS data, with the model capturing the elongation of the PDF (note however that the appearance of the elongation is somewhat exaggerated due to the different axis ranges) along the horizontal axis toward regions of large $Q_b/Q_{b,av}$, indicating that large fluctuations in the scalar gradients are much more probable than they are for the velocity gradients. The model also captures the exponential-like behaviour of the isocontours of the PDF, whose shape indicates that large values of $Q_b/Q_{b,av}$ tend to occur in regions where $Q/\tilde{Q}_{av}$ is small, and vice-versa. The quantitative errors in M1 are mainly associated with the variation of the joint-PDF along the $Q/\tilde{Q}_{av}$ axis, which can be understood in terms of the ML-RGDF's underprediction of large fluctuations of $Q/\tilde{Q}_{av}$ at $\Rey_\lambda=100$, as already observed when considering the PDF of $Q/\tilde{Q}_{av}$ in figure \ref{PDF_QR}. Comparing the results from M1, M2, and M3 shows that the model predicts that as $\Rey_\lambda$ is increased the shape of the joint-PDF is preserved, but becomes stretched along the two axes. Correspondingly the probability of regions with comparable values of $Q/\tilde{Q}_{av}$ and $Q_b/Q_{b,av}$ is predicted to decrease as $\Rey_\lambda$ increases.

{\vspace{0mm}\begin{figure}
		\centering
		\subfloat[]
		{\begin{overpic}
				[trim = 0mm 60mm 10mm 60mm,scale=0.3,clip,tics=20]{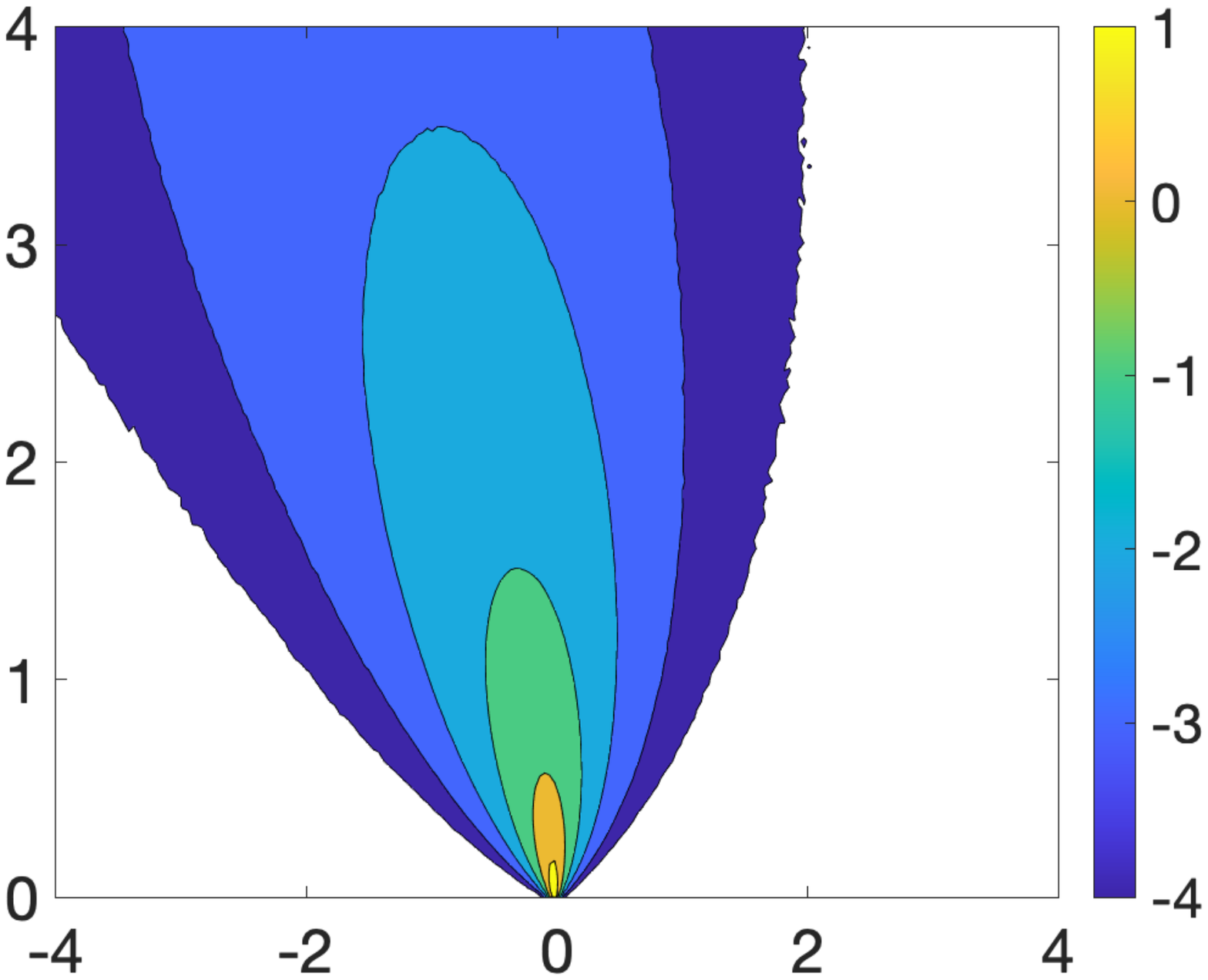}
				\put(70,-1){$R_b/\tilde{Q}_{av}^{1/2}Q_{b,av}$}
				\put(8,50){\rotatebox{90}{$Q_b/Q_{b,av}$}}					
		\end{overpic}}
		\centering
		\subfloat[]
		{\begin{overpic}
				[trim = 0mm 60mm 10mm 60mm,scale=0.3,clip,tics=20]{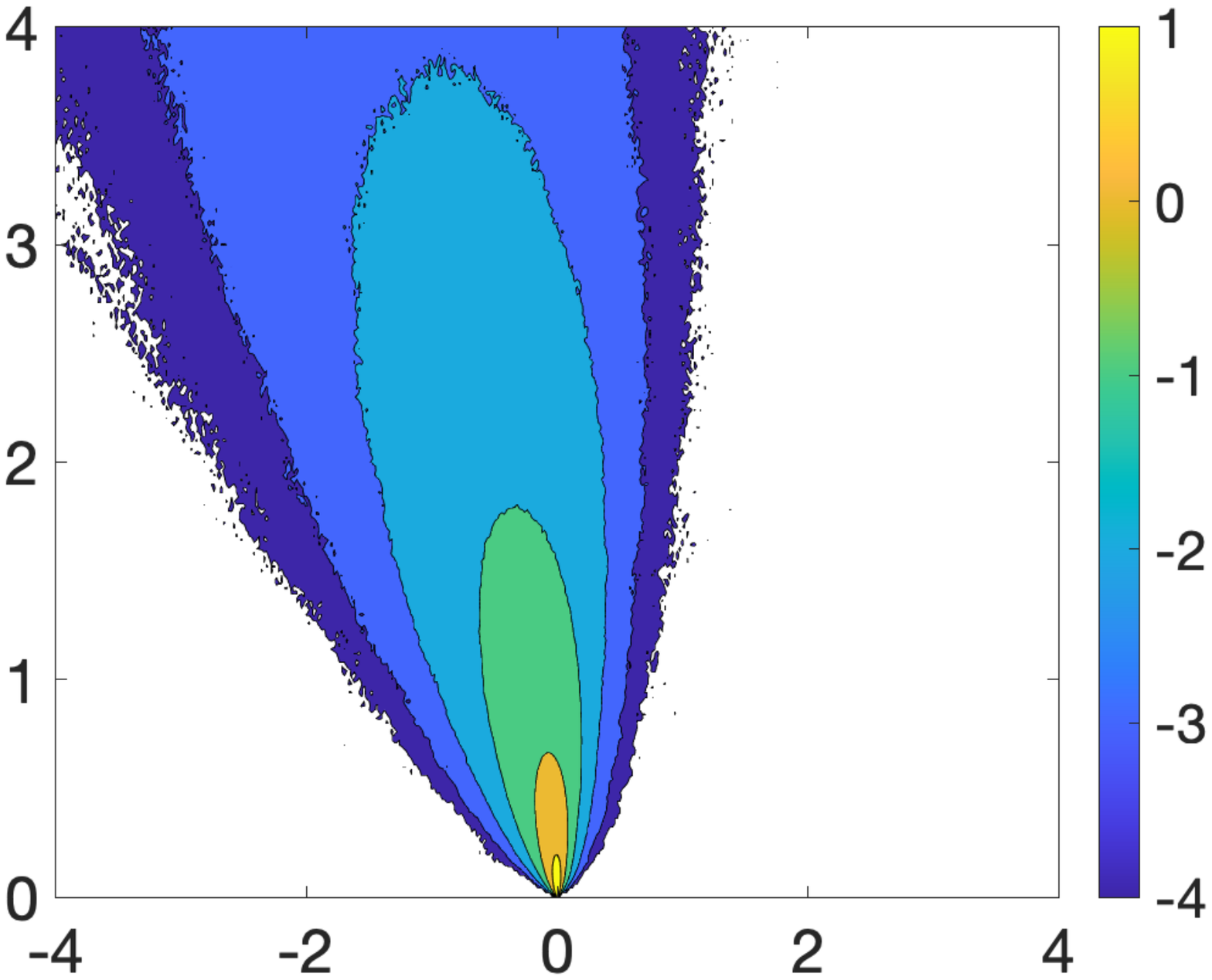}
				\put(70,-1){$R_b/\tilde{Q}_{av}^{1/2}Q_{b,av}$}
				\put(8,50){\rotatebox{90}{$Q_b/Q_{b,av}$}}			
		\end{overpic}}\\
		\centering
		\subfloat[]
		{\begin{overpic}
				[trim = 0mm 60mm 10mm 60mm,scale=0.3,clip,tics=20]{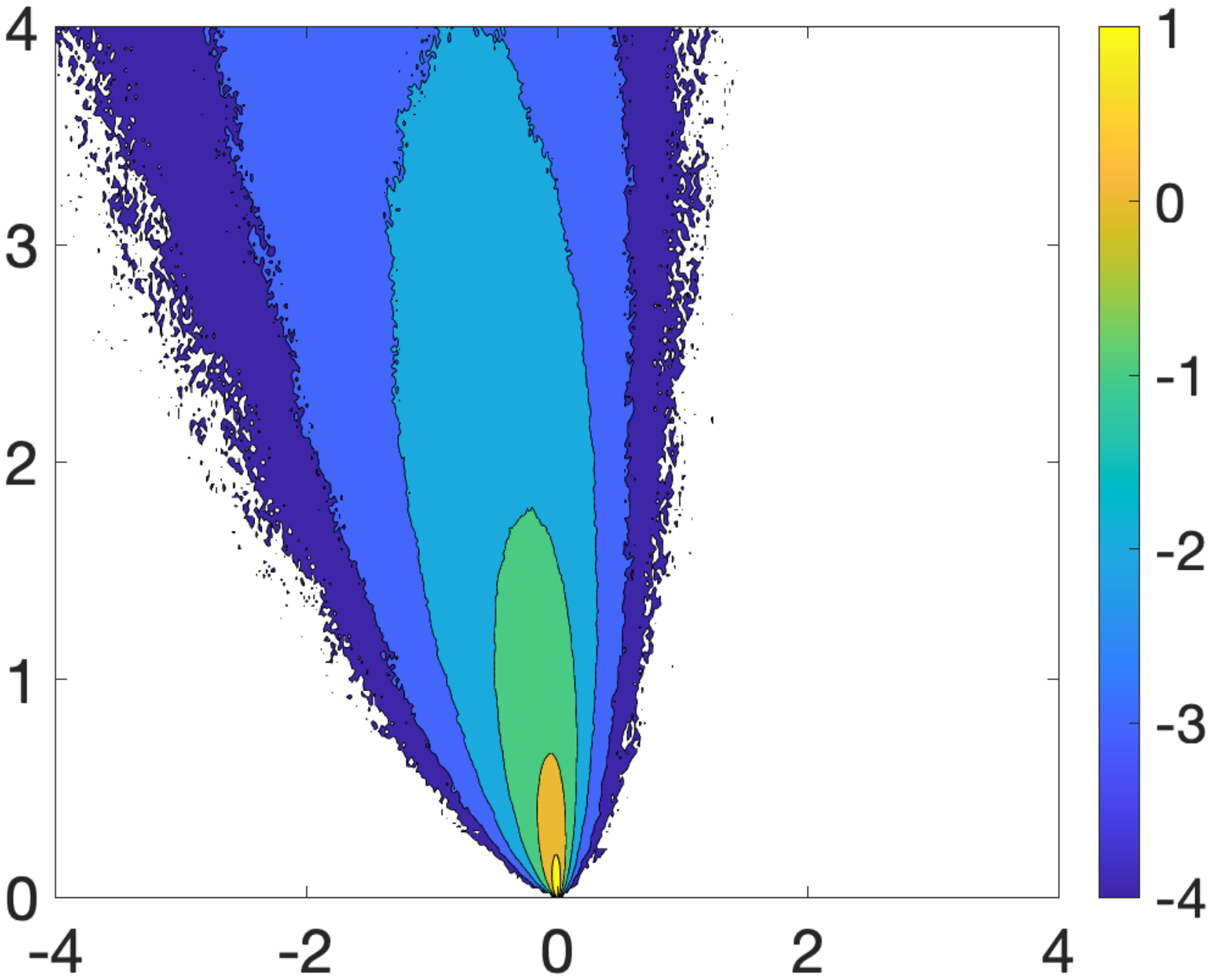}
				\put(70,-1){$R_b/\tilde{Q}_{av}^{1/2}Q_{b,av}$}
				\put(8,50){\rotatebox{90}{$Q_b/Q_{b,av}$}}					
		\end{overpic}}	
				\centering
		\subfloat[]
		{\begin{overpic}
				[trim = 0mm 60mm 10mm 60mm,scale=0.3,clip,tics=20]{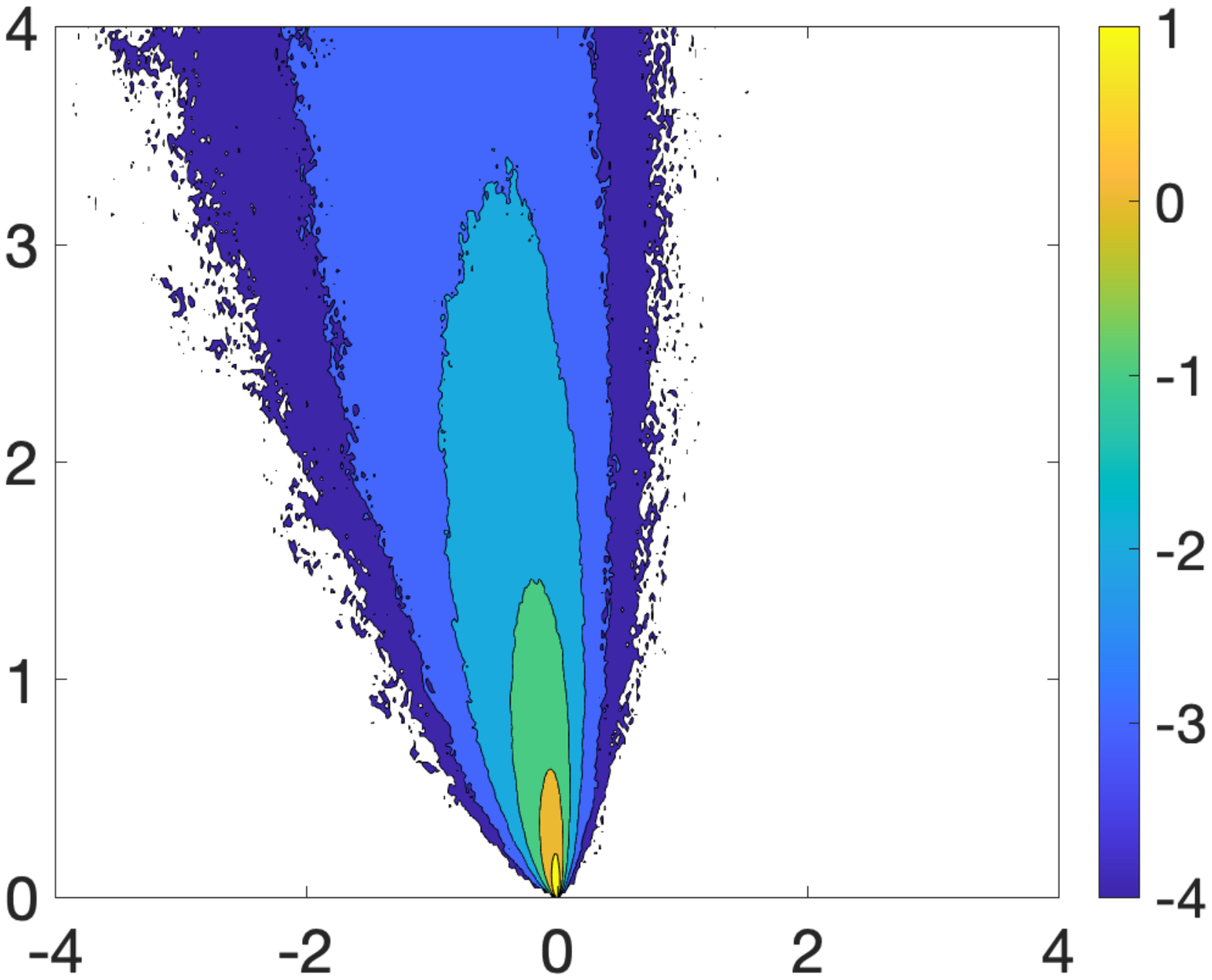}
				\put(70,-1){$R_b/\tilde{Q}_{av}^{1/2}Q_{b,av}$}
				\put(8,50){\rotatebox{90}{$Q_b/Q_{b,av}$}}				
		\end{overpic}}	
		\caption{Logarithm of joint-PDFs of $Q_b/Q_{b,av}$ and $R_b/\tilde{Q}_{av}^{1/2}Q_{b,av}$ from (a) DNS, (b) M1, (c) M2, (d) M3. Colors indicate the values of the logarithm of the PDF.} 
		\label{PDF_Qb_Rb}
\end{figure}}

In Figure \ref{PDF_Q_Rb} we show the joint-PDFs of the velocity gradient invariant $Q/\tilde{Q}_{av}$ and scalar production invariant $R_b/\tilde{Q}_{av}^{1/2}Q_{b,av}$ from the DNS and model. This joint PDF provides insights into how straining and vortical regions of the flow might contribute differently to the scalar gradient production. The model accurately reproduces the qualitative behaviour of the PDF seen in the DNS data, including the elongation  of the PDF along the $R_b/\tilde{Q}_{av}^{1/2}Q_{b,av}<0$ direction, associated with the predominance of scalar gradient production over destruction. The results also show that events with strong scalar gradient production are more probable in regions where $Q/\tilde{Q}_{av}<0$, despite the fact that $Q/\tilde{Q}_{av}$ itself has a positively skewed PDF. This is because the antisymmetric part of the velocity gradient, and therefore the vorticity, does not directly contribute to the invariant $R_b$, but only contributes indirectly through its impact on the local alignment of $\boldsymbol{\mathcal{B}}(t)$ with $\boldsymbol{\mathcal{S}}(t)$, whereas the strain-rate directly affects $R_b$. The model underpredicts the probability of the largest fluctuations, which again is principally due to the ML-RDGF underpredicting the intermittency of $Q/\tilde{Q}_{av}$. Comparing the results from M1, M2, and M3 shows that the model predicts that as $\Rey_\lambda$ is increased the shape of the joint-PDF is largely preserved, except for being stretched along the axes due to the increased intermittency of the flow. The model does seem to predict, however, that the overall probability of the system being in the quadrant $Q<0,R_b>0$ reduces as $\Rey_\lambda$ increases. Future comparisons with DNS at higher $\Rey_\lambda$ will be needed to assess the accuracy of this prediction. It is possible that this is a defect in the model that is in some way related to the failure of the model at higher $\Rey_\lambda$.

Finally, in Figure \ref{PDF_Qb_Rb} we show the joint-PDFs of the scalar gradient invariant $Q_b/Q_{b,av}$ and scalar production invariant $R_b/\tilde{Q}_{av}^{1/2}Q_{b,av}$. The prediction of the model for M1 is in very good agreement with the DNS, both qualitatively and quantitatively, with only small deviations. As with the other joint-PDFs, comparing the results from M1, M2, and M3 shows that the model predicts that as $\Rey_\lambda$ is increased the shape of the joint-PDF of $Q_b/Q_{b,av}$ and $R_b/\tilde{Q}_{av}^{1/2}Q_{b,av}$ is preserved, except for being stretched along the axes due to the increased intermittency of the flow.

\FloatBarrier

\section{Conclusions}

A Lagrangian model for passive scalar gradients in isotropic turbulence has been developed, with the scalar gradient diffusion term closed using the RDGF approach, which has recently been very successfully applied to close the equation for the fluid velocity gradients along fluid particle trajectories. This closure yields a diffusion term that is nonlinear in the velocity gradients, but linear in the scalar gradients,  and comparisons of the statistics generated by the closed model with DNS data revealed large errors. An investigation revealed that these large errors were due to the scalar gradient model generating erroneously large fluctuations, possibly due to the diffusion term being linear in the scalar gradient under the RDGF closure. This defect was addressed by incorporating into the closure approximation information regarding the scalar gradient production along the local trajectory history of the particle. With this modification, the closed form of the diffusion term is now a nonlinear functional of the scalar gradients, and the resulting model is in very good agreement with the DNS data.

Since the ML-RDGF model of \cite{johnson2017turbulence} is used to specify the velocity gradients in the scalar gradient equation, we begin by comparing its predictions with DNS data. In agreement with the results of \cite{johnson2017turbulence}, the model very accurately predicts the longitudinal and transverse components of the velocity gradients, and reproduces the key features of the joint-PDF of $Q$ and $R$, the second and third invariants of the velocity gradient tensor. However, a more quantitative test of the model predictions for the PDFs of $Q$ and $R$ individually against DNS data revealed that at least for $\Rey_\lambda=100$, the ML-RDGF significantly underpredicts the probability of large positive values of $Q$ and large negative values of $R$, which are primarily associated with regions of intense enstrophy and enstrophy production, respectively. These inaccuracies could then impact the accuracy of the scalar gradient model, and highlight the need for further improvements in the ML-RDGF model (although the model may predict the statistics of $Q$ and $R$ more accurately at higher $\Rey_\lambda$).

Comparisons between the scalar gradient model and DNS data for $\Rey_\lambda=100$ showed very good agreement. In particular, the model accurately predicts the squared magnitude of the scalar gradients (which are proportional to the scalar dissipation rates), as well as the individual components of the scalar gradients.  The model also captures well the PDF of the scalar production, including its strong negative skewess that is associated with the predominance of scalar gradient production of destruction, but slightly underpredicts the most extreme fluctuations of the scalar gradient production and destruction. Next, the PDFs of the inner product between the scalar gradient direction and the strain-rate eigendirections were considered, which provide insights into the nontrivial statistical geometry of the passive scalar and velocity gradient dynamics. The model is in excellent agreement with the DNS data regarding the strong preferential alignment between the scalar gradient and the compressional eigendirection. However, the model predicts a misalignment with the extensional eigendirection that is a little too strong, and a misalignment with the intermediate eigendirection that is a little too weak.

The ability of the model to capture the statistical relationship between the velocity and scalar dynamics was considered next, by considering various joint-PDFs of the velocity and scalar gradient tensors. The results showed excellent qualitative agreement with the DNS data, with some quantitative errors that seem to be rooted in the ML-RDGF model underpredicting the probability of extreme fluctuations in the $Q$ and $R$ invariants. The joint-PDF of the squared magnitude of the scalar gradient with the scalar gradient production term predicted by the model was in excellent qualitative as well as quantitative agreement with the DNS.

The predictions of the model at $\Rey_\lambda$ greater than that of the DNS were also considered, and the predictions are reasonable up to around $\Rey_\lambda\approx 500$. However, beyond this, the model breaks down and leads to extremely large scalar gradients that can even cause the numerical simulations of the model to blow up. Therefore, while the modification to the scalar gradient diffusion term that incorporates the scalar gradient production along the local trajectory history of the particle leads to excellent predictions from the model at lower $\Rey_\lambda$, it is not sufficient to prevent the model from generating extremely large fluctuations at high $\Rey_\lambda$ where intermittency in the velocity gradients can lead to very large local scalar gradient production events. Therefore, as anticipated earlier, developing an accurate model for scalar gradients in turbulence is in some ways more complicated than that for velocity gradients, because scalar gradient dynamics lack a mechanism similar to the pressure Hessian that controls the growth of the velocity gradients. For scalar gradients, the closure for the diffusion term is a delicate matter since this term alone is dynamically responsible for preventing finite-time singularities of the scalar gradients. A crucial point for future work is therefore to understand in more detail how the diffusion term regulates the growth of the scalar gradients, and developing a closure model that is sufficiently sophisticated to capture this.

Another aspect to be explored is the influence of the Schmidt number $Sc$ on the scalar gradients. While the model does capture a $Sc$ dependence in the scalar gradient diffusion term, given that the model breaks down for $Sc=1$ when $\Rey_\lambda>500$, it is likely that the model will also break down for $\Rey_\lambda<500$ when $Sc$ becomes sufficiently large, since both regimes promote the intensification of the scalar gradients.

\backsection[Acknowledgements]{This work used the Extreme Science and Engineering Discovery Environment (XSEDE), which is supported by NSF grant number ACI-1548562 \citep{xsede}. Specifically, the Expanse cluster was used under allocation CTS170009.}

\backsection[Funding]{A.D. Bragg and X. Zhang acknowledge support through a National Science Foundation (NSF) CAREER award \# 2042346}

\backsection[Declaration of interests]{The authors report no conflict of interest.}


\bibliographystyle{jfm}
\bibliography{jfm}

\end{document}